\documentclass[twocolumn,superscriptaddress,prl]{revtex4-2}
\usepackage{amsmath}
\usepackage{graphicx}
\usepackage{xcolor} 
\usepackage[normalem]{ulem} 

\newcommand{\stkout}[1]{\ifmmode\text{\sout{\ensuremath{#1}}}\else\sout{#1}\fi}
\newcommand{\md}{\mathrm d}

\newcommand{\br}{\boldsymbol{r}}
\newcommand{\bR}{\boldsymbol{R}}
\newcommand{\bp}{\boldsymbol{p}}
\newcommand{\bq}{\boldsymbol{q}}
\newcommand{\blambda}{\boldsymbol{\lambda}}

\begin{document}

\title{Delta-Function Kicks are Optimal for Rapidly Driven Inertial Stochastic Systems}

\author{Steven Blaber}
\email{steven.blaber@ubc.ca}
\affiliation{Dept.~of Physics and Astronomy and Stewart Blusson Quantum Matter Institute, University of British Columbia, Vancouver, British Columbia V6T 1Z1, Canada}
\begin{abstract}
Optimal control helps guide our understanding of stochastic thermodynamics, leading to universal properties and geometric formulations. Among the initially surprising properties of optimal control, not only discrete jumps but delta function kicks have been shown to be necessary to minimize dissipation in specific example systems. Using a short-time approximation, I show that delta-function kicks are universally optimal for minimizing dissipation in inertial stochastic systems, including active and quantum dynamics under general constraints. Fundamentally stemming from basic kinematics, delta-function kicks are required to achieve linear scaling of work with short protocol durations compared to the quadratic scaling without the kicks. This implies a diverging (infinite) ratio of saved work in the short-time limit.
\end{abstract}
\date{\today}

\maketitle
Among some of the foundational studies in both classical and stochastic thermodynamics is that of optimal control: how to perform a given task at minimal energetic cost~\cite{Carnot1960,andresen1977,salamon1977,andresen2011,Seifert2012,Brown2017,Brown2019}. Within stochastic thermodynamics, studies of optimal control resulted in the initially surprising development that discrete control parameter jumps are required to minimize dissipation in specific model systems~\cite{Schmiedl2007}, and later any stochastic or quantum thermodynamic system~\cite{Blaber2021,rolandi2023optimal}. When extending this to inertial systems, a further quirk was discovered, the optimal control of underdamped (inertial) systems with quadratic trapping potentials requires not only discrete jumps, but also delta-function kicks in the trap centre or stiffness at the start and end of the protocol~\cite{Gomez2008}. The improvement from optimal control compared to naive (linear) driving was observed to be significantly larger for underdamped compared to overdamped systems~\cite{Gomez2008}.

General solutions for optimal control in stochastic thermodynamics are typically difficult to determine; however, recent progress has been made towards exact solutions for general systems building on optimal-transport~\cite{zhong2022} and numerical techniques~\cite{Then2008,gingrich2016,engel2022}. While exact solutions are convenient when possible, the determination of minimum-dissipation protocols can be considerably simplified through approximate methods: linear-response theory for weak perturbations~\cite{kamizaki2022,bonanca2018}, thermodynamic-geometry~\cite{Sivak2016} for slowly driven systems under diverse conditions~\cite{Blaber2020,deffner2020,zulkowski2012,bonancca2014,zulkowski2015,zulkowski2015Quantum,Large2019,Lucero2019,Rotskoff2015,Rotskoff2017,Tafoya2019,louwerse2022,frim2022,frim2021}, and short-time efficient protocols (STEPs) for rapidly driven systems~\cite{Blaber2021}. 

Although many extensions have been made to more general dynamics~\cite{korbel2026quo}, optimal control theory in stochastic thermodynamics has primarily focused on overdamped dynamics. This is for good reason, typical biological molecular machines have negligible inertia. For example, Escherichia coli propelled by its flagellar motor through water will coast less than the width of an atom if it stops actively swimming~\cite{purcell2014life}. With this context in mind, it is still important to account for inertial effects in many other cases: molecular dynamics simulations and free energy estimation~\cite{Park2004}, levitated nanoparticles~\cite{baldovin2025optimal,rondin2017direct,militaru2021kovacs}, and micromechanical resonators~\cite{dago2021information,el2026extracting}. In any case, inertial dynamics are more fundamental and the overdamped scenario is a limiting case.

In this letter, I apply a short-time approximation to show that under very general conditions delta-function kicks are optimal for work minimizing protocols for inertial stochastic, quantum, or active systems, generalizing the original finding for harmonic potentials~\cite{Gomez2008}. The optimality of kicks fundamentally stems from the simple kinematic equation $x = x_0 + v_0t+\tfrac{1}{2}at^2$. Such universal results have profound implications for our understanding of thermodynamics, the design principles of molecular machines, and applicability of optimal control to engineered microscopic systems.

\textit{Klein-Kramers dynamics}---Consider inertial stochastic systems obeying dynamics that can be expressed as
\begin{align}
    \frac{\partial \Pi(\br,\bp,t)}{\partial t} = L[\br,\bp,\blambda(t)]\Pi(\br,\bp,t) \ , \label{eq: K-K operator}
\end{align}
for probability density $\Pi(\br,\bp,t)$, position $\bf r$, momentum $\bf p$, and time dependent control parameters $\boldsymbol{\lambda}(t)$. For underdamped dynamics, this takes the form of the Klein-Kramers equation
\begin{align}
    \frac{\partial \Pi}{\partial t}=&\xi \nabla_{\bp} \cdot\left(\bp \Pi\right)-\nabla_{\bp} \cdot(\boldsymbol{F}~ \Pi)-\frac{1}{m} \bp \cdot \nabla_{\br} \Pi \nonumber\\
    &+m \xi \beta^{-1} \nabla_{\bp}^2 \Pi \ , \label{eq: K-K}
\end{align}
with momentum damping rate $\xi$, mass $m$, and inverse temperature $\beta \equiv (k_{\rm B}T)^{-1}$. $\boldsymbol{F}[\br,\blambda(t)] = -\nabla_{\br}V[\br,\blambda(t)]$ is the conservative force applied to the system with energy V.

Multiplying Eq.~\eqref{eq: K-K} by $g[\br,\blambda(t)]$ and integrating over $\br$ and $\bp$ it follows that the average of any function $g[\br,\blambda(t)]$ that is independent of $\bp$ is
\begin{align}
    \frac{\partial\langle g\rangle_{\Lambda}}{\partial t} = \frac{1}{m}\langle \bp\cdot\nabla_{\br} g \rangle_{\Lambda} \ ,
    \label{eq: Average of p ind. variable}
\end{align}
where $\partial/\partial t$ acts only on the probability distribution through $L$ and assuming the boundary terms vanish. Angle brackets $\langle\cdots\rangle_{\Lambda}$ denote the nonequilibrium average over $\Pi(\br,\bp,t)$ given a control protocol $\Lambda$. For short protocol duration $\Delta t$ we expand the probability distribution to first order as~\cite{Blaber2021}
\begin{align}
    \Pi(\br,\bp,t) \approx \Pi(\br,\bp,0) + \int_{0}^{t}\md t' L[\br,\bp,\blambda(t')]\Pi(\br,\bp,0) \ , \label{eq: K-K approx}
\end{align}
implying
\begin{align}
     \langle g\rangle_{\Lambda} \approx \langle g\rangle_{\blambda_0} + \frac{1}{m}\int_{0}^{t}\md t' \left\langle \bp\cdot\nabla_{\br} g \right\rangle_{\blambda_0} \ .
    \label{eq: short-time Average of p ind. variable}
\end{align}
Throughout we use $\langle \cdots\rangle_{\blambda}$ to denote an equilibrium average at $\blambda$, while $\Lambda$ is the nonequilibrium average. We define $\blambda_0$ and $\blambda_{\rm f}$ as the (fixed) initial and final control parameter values respectively.

The quantity we wish to optimize is the total average work done by time dependent control of $\blambda$
\begin{align}
    \langle W \rangle_{\Lambda} = -\int_{0}^{\Delta t}\md t \frac{\md\blambda(t)}{\md t}\cdot \langle \boldsymbol{f}(t) \rangle_{\Lambda} \ ,
\end{align}
where we have defined the conjugate force $\boldsymbol{f}\equiv -\partial V/\partial\blambda $. For optimization in the short-time limit, it is useful to consider the \emph{saved work}, the difference in work from an instantaneous protocol
\begin{align}
    \langle W_{\rm saved} \rangle_{\Lambda} \equiv \langle W \rangle_{\blambda_{0}} -\langle W \rangle_{\Lambda} \ .
    \label{eq: Saved work definition}
\end{align}
Minimizing work is equivalent to maximizing the saved work.

The only unknown quantity for optimization in \eqref{eq: Saved work definition} is $\langle\boldsymbol{f}(t) \rangle_{\Lambda}$. Applying the short-time approximation in Eq.~\eqref{eq: K-K approx}, the average conjugate force is
\begin{align}
    \langle\boldsymbol{f}(t) \rangle_{\Lambda} \approx \langle\boldsymbol{f} \rangle_{\blambda_{0}} + \int_{0}^{t}\md t'\bR_{\boldsymbol{f}}[\blambda(t')] \ ,
\end{align}
with $\bR_{\boldsymbol{f}}$ the initial force-relaxation rate, defined as
\begin{align}
    \bR_{\boldsymbol{f}}[\blambda(t)] \equiv \int\int\md\br\md\bp~ \boldsymbol{f}[\br,\blambda(t)] L[\br,\bp,\lambda(t)]\Pi(\br,\bp,0) \ .
\end{align}
Substituting into Eq.~\eqref{eq: Saved work definition} and integrating by parts, the saved work is~\cite{Blaber2021}
\begin{align}
\langle W_{\rm saved} \rangle_{\Lambda} \approx \int_{0}^{\Delta t}\md t  \, \bR_{\boldsymbol{f}}[\boldsymbol{\lambda}(t)]\cdot
[\boldsymbol{\lambda}_{{\rm f}} - \boldsymbol{\lambda}(t)] \ .
\end{align}

At first glance, this appears optimized by a specific set of control parameters $\blambda^{\rm STEP}$, implying jumps to and from these values at both ends of the protocol just as was the case for overdamped dynamics; however, for such a protocol the initial force relaxation rate $\boldsymbol{R}$ is zero. This is because, assuming the system is initially at equilibrium and the potential is independent of momentum, the conjugate force $\boldsymbol{f}$ is independent of $\bp$. Substituting in Eq.~\eqref{eq: short-time Average of p ind. variable} we have 
\begin{align}
    \bR_{\boldsymbol{f}} \propto \left\langle \bp\cdot \nabla_{\br}\boldsymbol{f}\right\rangle_{\lambda_{0}}
\end{align}
Since it is initially at equilibrium, $\br$ and $\bp$ are uncorrelated and $\langle \bp \rangle_{\blambda_{0}}=0$, so the initial-force relaxation rate is always zero. In this case, the $\mathcal{O}(\Delta t)$ term is zero and the saved work would scale as $\langle W_{\rm saved} \rangle_{\Lambda} \sim (\Delta t)^2$ rather than $\langle W_{\rm saved} \rangle_{\Lambda} \sim  \Delta t$ as was the case for overdamped dynamics.

There are only two ways to achieve the faster short-time $\mathcal{O}(\Delta t)$ dynamical scaling: either by introducing a momentum dependent potential such that $\boldsymbol{f} = \boldsymbol{f}[\br,\bp,\blambda(t)]$ or by introducing delta function impulses that deterministically changes the initial momentum of the system such that $\left\langle \bp\cdot \nabla_{\br} \boldsymbol{f}\right\rangle_{\lambda_{0}} \neq 0$. An initial non-zero average momentum can be achieved by applying a delta function in the force at $t=0$. Inducing correlations between position and momentum is achieved by making this perturbation position dependent.

Applying a delta function in the force at $t = 0$ and $t=\Delta t$ we have $\boldsymbol{F}[\br,\blambda(t)] + \boldsymbol{A}(\br)\delta(t) - \boldsymbol{B}(\br)\delta(t-\Delta t)$. Integrating Eq.~\eqref{eq: K-K} over the infinitesimal time interval of the delta function impulse reveals that it deterministically shifts the momentum 
\begin{align}
    \Pi(\br,\bp,0^+) = \Pi(\br,\bp-\boldsymbol{A}(\br),0) \ ,
    \label{eq: Shifted probability}
\end{align}
where we have assumed $\boldsymbol{A}$ is independent of $\bp$.

The work done by these deterministic shifts in momentum is equal to the change in kinetic energy of the system. Since the initial and final impulses shift the momentum as $\bp\to\bp+\boldsymbol{A}$ and $\bp\to\boldsymbol{p}-\boldsymbol{B}$ respectively, the total change in kinetic energy from these two impulses is
\begin{align}
    \langle W^{\delta} \rangle &= \frac{\langle(\bp+\boldsymbol{A})^2-\bp^2\rangle_{\blambda_{0}}}{2m} + \frac{\langle(\bp-\boldsymbol{B})^2-\bp^2\rangle_{\Lambda}}{2m}\\
    &= \frac{\langle\boldsymbol{A}^2\rangle_{\blambda_{0}}}{2m} + \frac{\langle\boldsymbol{B}^2\rangle_{\Lambda}}{2m} - \frac{\langle\bp\cdot\boldsymbol{B}\rangle_{\Lambda}}{m} \ .
    \label{eq: impulse work}
\end{align}
Where in the second line we assumed the initial average momentum is $0$ and uncorrelated with $\boldsymbol{A}$.

Expanding the impulses in $\Delta t$ as $\boldsymbol{A} \approx \boldsymbol{A}_{0} + \boldsymbol{A}_{1}\Delta t + \dots$ we have
\begin{align}
    \langle W^{\delta}  \rangle \approx \frac{1}{2m}\left\langle \boldsymbol{A}^2_0+\boldsymbol{B}^2_0-2\bp\cdot\boldsymbol{B}_0\right\rangle_{\blambda_{0},\bp^+} \ ,
\end{align}
with the average $\langle\cdots\rangle_{\blambda_{0},\bp^+}$ is defined to be an average over the initial distribution after the impulse $\Pi(\br,\bp-\boldsymbol{A}(\br),0)$. The leading order terms, which are independent of $\Delta t$, can be eliminated by setting $\boldsymbol{B}_{0} = \boldsymbol{A}_{0}$. This helps to clarify the need for the second impulse. To $0^{\rm th}$ order, energy is injected into the system adiabatically in the form of kinetic energy at the initial impulse which must then be extracted in an identical manner at the end.  Accounting for the dynamical evolution of the system, the final impulse must be adjusted and will not be symmetric in general. The total saved work with impulses is 
\begin{align}
    \langle W_{\rm save} \rangle_{\Lambda} \approx \int_{0}^{\Delta t}\md t  \, \bR^{0^+}_{\boldsymbol{f}}[\boldsymbol{\lambda}(t)]\cdot
[\boldsymbol{\lambda}_{{\rm f}} - \boldsymbol{\lambda}(t)] - \langle W^{\delta} \rangle \ ,
\end{align}
where the initial force relaxation rate is evaluated after the initial impulse (denoted by superscript $0^+$).

One very interesting consequence of this analysis is the scaling of the saved work. Comparing the optimal to naive impulse free work we have
\begin{align}
    \frac{\langle W_{\rm saved}^{\rm optimal}\rangle}{\langle W_{\rm saved}^{\rm naive}\rangle} \sim \frac{1}{\Delta t} \ ,
\end{align}
for small $\Delta t$. This results in an infinitely large ratio of work savings compared to naive driving as $\Delta t \rightarrow 0$. For comparison, a quadratic trapping potential with overdamped dynamics has a constant optimal to naive saved work ratio of $3/2$. This explains the much larger improvement from optimal control observed for harmonic trapping potential in underdamped compared to overdamped dynamics~\cite{Gomez2008}.

The linear versus quadratic scaling in work and dynamics for inertial systems has a very simple physical and mathematical interpretation. Consider the basic kinematic equation from classical dynamics
\begin{align} \label{eq: classical kinematics}
    x(t) = x_{0} + v_0 t +\frac{1}{2}a t^2 \ .
\end{align}
Evidently, for zero initial velocity $v_0 = 0$ the position evolves as $x(t)\sim t^2$, while if we have $v_0\neq 0$ we have $x(t)\sim t$ which dominates for short duration protocols. This analogy is made exact by expanding inertial stochastic dynamics for short-times as
\begin{align}\label{eq: stochastic kinematic}
    \langle x(t)\rangle \approx\langle x \rangle_0 + \langle v \rangle_{0}t +\frac{1}{2}\langle a\rangle_{0}t^2 \ .
\end{align}
For zero initial average velocity (momentum) $\langle v \rangle_{0}=0$, the dynamics evolves as $\langle x(t)\rangle\sim t^2$ while for $\langle v \rangle_{0}\neq0$ we have $\langle x(t)\rangle\sim t$  which dominates for short times. A similar expansion holds for any function $g(x)$ that is independent of momentum, giving Eq.~\eqref{eq: short-time Average of p ind. variable}. Fundamentally, the delta-function kicks are necessary to minimize work due to basic kinematics and the optimality of delta-function kicks should be the expected result while the absence of kicks for overdamped dynamics is an artifact of strong damping breaking down classical kinematics.

The same mechanism extends beyond Markovian Klein--Kramers dynamics whenever the state evolution satisfies Eq.~\eqref{eq: Average of p ind. variable} for momentum-independent observables and admits impulsive momentum shifts. This includes generalized Langevin dynamics with momentum-space noise (active systems) and quantum Brownian motion when the dissipator does not act directly on position (Supplemental Material~\ref{section: Active and Quantum}). Beyond inertial systems, stochastic RLC circuits obey analogous dynamics~\cite{sabbagh2024wasserstein} and will also have optimal delta function kicks for work~\cite{freitas2020stochastic} minimizing protocols under inductance control $L(t)$, but not capacitive control $C(t)$ since capacitance is mass-like which would alter the analog of momentum. The optimality of impulses was observed numerically for quantum non-Markovian dynamics in an impulse-Ansatz approach~\cite{tokieda2025work}. Derivations and the limitations imposed by direct configurational noise and completely positive Lindblad corrections are given in Supplemental Material~\ref{section: Active and Quantum}.

\textit{Universal short-time kicks}---Now that we have established that delta-function kicks are universally optimal for minimizing work in inertial stochastic systems, we turn our attention to the determination of the kick magnitude in general systems resulting in simple universal impulses. We will refer to a protocol consisting of both initial/final ($\boldsymbol{A}$/$\boldsymbol{B}$) delta function kicks and initial/final discrete control parameter jumps with the control parameters held constant at the short-time optimal value $\blambda^*$ in between as a $\delta$-STEP ($\delta$-Short-Time Efficient Protocol).

The total work of a $\delta$-STEP can be separated into potential (pot) and kinetic ($\delta$) contributions
\begin{align}
    \langle W \rangle_{\Lambda} = \langle W^{\rm pot} \rangle_{\Lambda} + \langle W^{\delta} \rangle_{\Lambda} \ .
\end{align}
Since the process consists of two discrete jumps, the potential energy work is expressed in terms of changes in potential energy
\begin{align}
    \langle W^{\rm pot} \rangle_{\Lambda} = \langle V_{\blambda^*} - V_{\blambda_{0}} \rangle_{\blambda_{0},\bp^+}+\langle V_{\blambda_{\rm f}}  -  V_{\blambda^*} \rangle_{\Lambda} \ .
\end{align}
With short-hand notation $V_{\blambda} \equiv V(\br,\blambda)$. The difference in work from an instantaneous protocol (saved work) is
\begin{align}
   \langle W^{\rm pot}_{\rm saved} \rangle_{\Lambda} = \langle V_{\blambda_{\rm f}}-V_{\blambda^*} \rangle_{\blambda_{0},\bp^+} -\langle V_{\blambda_{\rm f}}- V_{\blambda^*} \rangle_{\Lambda} \ .
\end{align}
Applying Eq.~\eqref{eq: short-time Average of p ind. variable} to the nonequilibrium probabilities in the averages $\langle \cdots\rangle_{\Lambda}$ we have
\begin{align}
    \langle W^{\rm pot}_{\rm saved} \rangle_{\Lambda} \approx \frac{\Delta t}{m}\left\langle \boldsymbol{A}\cdot\left( \boldsymbol{F}_{\blambda_{\rm f}}-\boldsymbol{F}_{\blambda^*} \right) \right\rangle_{\blambda_{0}} \ .
    \label{eq: potential work}
\end{align}

Applying Eqs.~\eqref{eq: Average of p ind. variable}-\eqref{eq: short-time Average of p ind. variable} to the work done by the impulses (Eq.~\eqref{eq: impulse work}) is slightly more involved than the potential energy work since the last average involves momentum so~\eqref{eq: Average of p ind. variable} does not apply to this term. In addition, $\boldsymbol{B}$ must be chosen to exactly cancel the excess momentum remaining from $\boldsymbol{A}$ at the end of the protocol. Nevertheless, straightforward algebra (Supplemental Material~\ref{App: Impulse work derivation}) reveals a surprisingly simple form
\begin{align}
    \langle W^{\delta} \rangle_{\Lambda} \approx \frac{\Delta t}{m}\left\langle\xi\boldsymbol{A}^2 - \boldsymbol{A}\cdot\left( \boldsymbol{F}_{\blambda^*}-\boldsymbol{F}_{\blambda_0} \right)\right\rangle_{\blambda_{0}} \ .
    \label{eq: Kinetic work}
\end{align}
Combining Eqs.~\eqref{eq: potential work} and~\eqref{eq: Kinetic work} and defining $\Delta \boldsymbol{F}\equiv \boldsymbol{F}_{\blambda_{\rm f}}-\boldsymbol{F}_{\blambda_0}$, we have the total saved work for a $\delta$-STEP 
\begin{align}
    \langle W_{\rm saved} \rangle_{\Lambda} \approx \frac{\Delta t}{m}\left[ \left\langle \boldsymbol{A}\cdot\Delta \boldsymbol{F}\right\rangle_{\blambda_{0}} - \xi\langle\boldsymbol{A}^2\rangle_{\blambda_{0}}\right] \ .
\end{align}

The saved work is maximized up to order $\Delta t$ by
\begin{align}\label{eq: Optimal short-time impulse}
    \boldsymbol{A}^* &= \frac{\Delta \boldsymbol{F}}{2\xi} \ ,
\end{align}
and $\boldsymbol{B}^* =\boldsymbol{A}^*$ to leading order, resulting in saved work
\begin{align}
    \langle W^{\delta{\rm -STEP}}_{\rm saved}\rangle_{\Lambda} \approx \frac{\Delta t}{4m\xi}\left\langle|\Delta \boldsymbol{F}(\br)|^2 \right\rangle_{\blambda_{0}} \ .
\end{align}
Higher order corrections to $\boldsymbol{A}^*$ and $\boldsymbol{B}^*$ depend on the optimal STEP values $\blambda^*$, which are determined by expanding the saved work to at least $\mathcal{O}(\Delta t^2)$ and will depend on system specific details. In general, the delta-function kicks the system in the direction of the change in force, dampened by a factor of $2\xi$ due to the quadratic cost of moving through a viscous fluid. The final impulse, $\boldsymbol{B}^*$ is chosen to cancel the excess momentum from the initial kick remaining at the end of the protocol, adiabatically retrieving some of the input kinetic energy.

For affine parametric control, the force is $\boldsymbol{F} = \boldsymbol{F}_0+\sum_{\alpha}\lambda^{\alpha}\boldsymbol{G}_{\alpha}(\br)$ with linear coupling between control parameter $\lambda^\alpha$ to the force through $\boldsymbol{G}_{\alpha}$. In this case, to minimize work to order $\mathcal{O}(\Delta t)$, the optimal impulse control is implemented directly as
\begin{align}
    \blambda^{\delta} = \frac{\Delta\blambda}{2\xi}\left[\delta(t) -\delta(t-\Delta t)\right] \ .
    \label{eq: Affine control}
\end{align}

\textit{Driven Barrier Crossing}---In Supplemental Material~\ref{App: Quadratic} we solve for the optimal control of a driven quadratic trapping potential in an arbitrary energy landscape. In this section, we consider a fixed-stiffness quadratic trap translated through an arbitrary static landscape, $\boldsymbol{F} = K [\br_{\rm trap}(t) - \br] - \nabla_{\br}V_{\rm land}(\br )$ so Eq.~\eqref{eq: Affine control} gives 
\begin{align}\label{eq: Optimal quadratic trap centre short-time impulse 1}
    \br_{\rm trap}^\delta = \frac{\Delta\br_{\rm trap}}{2\xi}\left[\delta(t) -\delta(t-\Delta t)\right] \ .
\end{align}
These optimal kicks are in agreement with the original harmonic trapping potential protocol from Ref.~\cite{Gomez2008} taken to leading order; however, our result is valid for arbitrary static energy landscapes while the original derivation is only valid for harmonic potentials in isolation.

We simulate underdamped dynamics for driven barrier crossing with a quadratic trapping potential across a double well energy landscape
\begin{align}
    V[x,\lambda(t)] = \frac{k}{2}\left[x-\lambda(t) \right]^2 +V_{\rm B}\left[\left(\frac{x}{a}\right)^2-1\right]^2 \ ,
\end{align}
with energy barrier $V_{\rm B}$ and symmetric minima at $x=\pm a$. We compare three different protocols, $\delta-$STEP described by Eq.~\eqref{eq: Optimal quadratic trap centre short-time impulse}, $\delta$-linear protocol which has the impulses as described by Eq.~\eqref{eq: Optimal quadratic trap centre short-time impulse 1} but with a linear trap center protocol, and a linear protocol with no impulses or control parameter jumps. We include the $\delta$-linear protocol since it is known that the optimal protocol for long protocol duration is continuous and the STEP performs poorly in this limit~\cite{Gomez2008,Blaber2021}. The $\delta$-function impulses in this protocol are $A^{\delta{\rm -linear}}=A^*/(1+\Delta t/\tau)$ with $\tau \equiv m\xi/k$ the position-position autocorrelation relaxation time.

\begin{figure}
	\includegraphics[width=\linewidth]{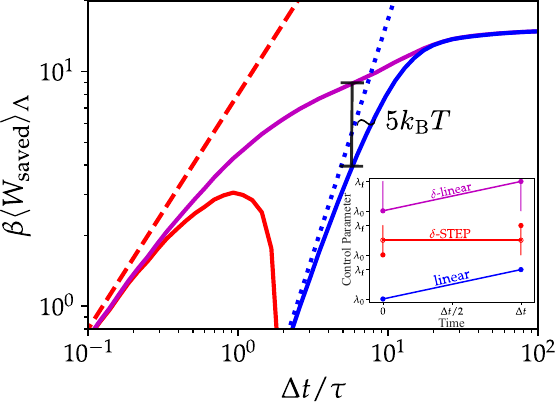}
    \caption{Saved work three different protocols. Dashed and dotted lines are the short time approximations, scaling as $\sim\Delta t$ and $\sim (\Delta t)^2$ respectively. Inset: Schematic diagram of the three protocols implemented: a linear protocol (blue), $\delta-$STEP (red) described by Eq.~\eqref{eq: Optimal quadratic trap centre short-time impulse}, and $\delta$-linear (purple) protocol which has the impulses as described by Eq.~\eqref{eq: Optimal quadratic trap centre short-time impulse 1} but with a linear trap center protocol. The energy scale is set by  thermal energy $k_{\rm B} T$ and time by the systems natural relaxation time $\tau \equiv m\xi/k$. All simulations have $\beta V_{\rm B} = 1$, $ka^2/V_{\rm B}=8$, $k/(m\xi^2) =8$, $\lambda_0/a = -1$, and $\lambda_{\rm f}/a = 1$.}
	\label{fig: saved work}
\end{figure}

Comparing to the linear protocol in Fig.~\ref{fig: saved work}, for short duration ($\Delta t/\tau \ll 1$) both the $\delta$-STEP and $\delta$-linear significantly outperform the kick free linear protocol. The $\delta$-linear and $\delta$-STEP perform similarly well in this limit, both scaling as $\sim\Delta t$ with identical coefficients while the impulse-free linear protocol scales as $\sim(\Delta t)^2$. The identical scaling of $\delta$-STEP and $\delta$-linear is because the STEP correction is of order $\mathcal{O}(\Delta t)^2$. For longer duration ($\Delta t/\tau \gg 1$), the $\delta$-STEP performs worse than the linear since holding $\blambda$ fixed is a bad strategy for long durations~\cite{Blaber2021}; however, the $\delta$-linear can save over $5k_{\rm B}T$ when traversing a $1k_{\rm B} T$ barrier at intermediate protocol duration and continues to perform well at long protocol durations. This large energy saving stems from the $\mathcal{O}(\Delta t)$ difference in work scalings, highlighted for short duration $\Delta t/\tau \ll 1$.

\textit{Discussion}---I have shown that delta-function impulses are necessary to optimally control inertial stochastic thermodynamic systems under specific general constraints. The conditions that must be satisfied are: i) the system begins in equilibrium with zero average momentum that is uncorrelated from the spatial coordinates $\br$, ii) applied forces are independent of momentum and iii) the dynamics must be inertial $\md \br/\md t = \bp/m$ with no direct spatial noise. Stemming from basic kinematics, the impulses are required in order to achieve $\sim\Delta t$ work scaling in the short time limit, resulting in a diverging work savings ratio for optimal compared to naive impulse free control for short protocol durations. The leading order impulse for general dynamics takes on the strikingly simple form, Eqs.~\eqref{eq: Optimal short-time impulse}-\eqref{eq: Affine control}. This considerably improves our understanding of stochastic thermodynamic systems.

The most direct application of the results derived in this article are to optically levitated nano-particles and steered molecular dynamics simulations. For the former, delta-function kicks cannot be directly implemented due to physical constraints; however, they can be approximated with strong kicks constrained to maximum force values. Optimal control strategies involving these more physical variations of delta-function kicks have been shown to reduce dissipation in optically levitated nanoparticles~\cite{baldovin2025optimal}. Simulated systems allow for more direct implementation of arbitrary control protocols without the same physical constraints, and it has been shown that the accuracy of free energy estimates is improved by protocols that minimize dissipation~\cite{Shenfeld2009,Blaber2020Skewed}. Given that the delta-function kicks not only reduce dissipation, but are necessary to achieve $\sim\mathcal{O}(\Delta t)$ scaling of dynamics and dissipation compared to the $\sim\mathcal{O}(\Delta t^2)$ without the kicks, they may be able to significantly reduce dissipation and improve free energy estimates from rapid driving protocols.

One exciting area of application is to generative modeling. For overdamped dynamics, there is a clear connection between diffusion modeling and optimal control in stochastic thermodynamics. Speed-accuracy relations for diffusion models show that the accuracy of data generation is constrained by entropy production (dissipation) with the optimal learning protocol given by optimal transport theory~\cite{ikeda2025speed}. Although overdamped diffusion models remain the predominant generative algorithm, there is evidence that underdamped noising/denoising processes can improve the algorithm's performance~\cite{dockhorn2021score,blessing2025underdamped}; however, optimal transport for underdamped dynamics can be significantly more complex than overdamped~\cite{dechant2019,sabbagh2024wasserstein}. It will be interesting to see if general optimal control principles in underdamped dynamics can be applied to improve generative modeling algorithms.

\section*{Acknowledgments}
I thank Masaaki Tokieda, and Sarah A.M. Loos for insightful discussions that inspired this work, Jörg Rottler for many discussions of this work and critical reading of the manuscript. Finally, I thank the Yukawa Institute for Theoretical Physics at Kyoto University for the ``Kyoto Workshop on Quantum Thermodynamics and Stochastic Thermodynamics 2025" (YITP-I-25-03), supported by MEXT KAKENHI Grant-in-Aid for Transformative Research Area (B) ``Quantum Energy Innovation" (Grant Number 24H00830) which helped inspire this work. This research was supported in part by a Natural Sciences and Engineering Research Council of Canada (NSERC) Canada Postdoctoral Research Award (CPRA).

\setcounter{section}{0}

\onecolumngrid
\clearpage
\begin{center}
	\textbf{\large Supplemental Material}
\end{center}
\setcounter{equation}{0}
\setcounter{figure}{0}
\setcounter{table}{0}
\setcounter{page}{1}
\makeatletter
\renewcommand{\theequation}{S\arabic{equation}}
\renewcommand{\thefigure}{S\arabic{figure}}

\section{Active and quantum systems}\label{section: Active and Quantum}

The analysis presented in the previous section applies to any dynamics of the form~\eqref{eq: K-K operator} that yield a statistical average like \eqref{eq: Average of p ind. variable} that can be shifted by a delta function kick as \eqref{eq: Shifted probability}. For a system described by the generalized Langevin equation with arbitrary memory kernel $\Gamma$ and colored noise $\eta_{a}$ we have~\cite{adelman1976fokker}
\begin{align}\label{eq:active_GLE}
\frac{\md\br(t)}{\md t} &= \frac{\bp(t)}{m} \ , \\
\frac{\md\bp(t)}{\md t} &= \boldsymbol{F}[\br(t),\blambda(t)]
-\int_0^t \md t'~\Gamma(t-t')\frac{\bp(t')}{m} + \eta(t) + \eta_{\rm a}(t) \ . \nonumber
\end{align}
For any function $g(\br)$ independent of $\bp$ we have
\begin{align}
    \md g(\br,\blambda) = \nabla_{\br}g(\br,\blambda)\cdot\md\br \ .
\end{align}
The explicit time-dependence through the full derivative $[\md \blambda(t)/\md t]\cdot [\partial g/\partial \blambda(t)]$ is deterministic and does not contribute to energy dissipation. 
Substituting in Eq.~\eqref{eq:active_GLE} and dividing by $\md t$ gives
\begin{align}
    \frac{\md g(\br)}{\md t} = \frac{1}{m}\bp\cdot\nabla_{\br}g(\br) \ .
\end{align}
Averaging over many trajectories yields the identical form as Eq.~\eqref{eq: Average of p ind. variable}
\begin{align}
    \frac{\partial \langle g\rangle_{\Lambda}}{\partial t} = \frac{1}{m}\langle \bp\cdot\nabla_{\br} g \rangle_{\Lambda} \ ,
\end{align}
and a momentum impulse is required to minimize work in the short-time limit. In general, the specific optimal values of the kicks will depend on the details of the active noise and memory kernel; most obviously through a change in the amount of viscous damping. If the active dynamics contains direct configurational noise or self-propulsion in the position equation then these relations break down.

We define the average work for quantum systems as~\cite{rolandi2023optimal},
\begin{align}
    \langle W \rangle_{\Lambda}=\int_0^{\Delta t} \md t\,\dot{\blambda}(t)\left\langle\frac{\partial {H}[{\br},{\bp},\blambda(t)]}{\partial\blambda}\right\rangle \ ,
    \label{eq: quantum work}
\end{align}
where the average represents a trace over the density matrix $\rho$, and ${H}$ is the Hamiltonian operator
\begin{align}
    {H}[{\br},{\bp},\blambda(t)] = \frac{{\bp}^2}{2m}+{V}[{\br},{\bp},\blambda(t)] \ .
\end{align}
The density matrix evolves as~\cite{isar1994open}
\begin{align}
    \frac{\md{\rho}}{\md t}  = L[{\br},{\bp},\blambda(t)] \rho \ ,
\end{align}
implying that
\begin{align}
    \frac{\md \langle g(\br,\blambda)\rangle}{\md t} = \langle L^{\dagger}g(\br,\blambda)\rangle \ ,
\end{align}
for operator adjoint $\dagger$. The time evolution operator satisfies 
\begin{align}
    L^{\dagger}O = \frac{i}{\hbar}[H,O]+\mathcal{D}^{\dagger}O \ .
\end{align}
The first term is the usual isolated dynamical evolution of the system expressed in terms of the commutator $[\cdots,\cdots]$. The second term accounts for system-bath coupling through the dissipator $\mathcal{D}$. We assume semi-classical quantum Brownian motion where the dissipator does not directly couple to position, implying $\mathcal{D}^{\dagger}g = 0$. In addition, we have $[V,g] = 0$ since they are both momentum independent. This implies
\begin{align}
    L^{\dagger}g(\br,\blambda) &= \frac{i}{2m\hbar}[\bp^2,g(\br,\blambda)] \\
    L^{\dagger}g(\br,\blambda) &= \frac{1}{2m}\{\bp,\nabla_{\br}g(\br,\blambda)\} \ ,
\end{align}
for the anticommutator $\{\cdots,\cdots\}$. Averaging over the density matrix we have
\begin{align}\label{eq: Quantum g dynamics}
    \frac{\md \langle g(\br,\blambda)\rangle}{\md t} = \frac{1}{2m}\langle\{\bp,\nabla_{\br}g(\br,\blambda)\}\rangle \ .
\end{align}
Following identical steps as outlined in Section~\emph{Klein-Kramers Dynamics}, it follows that Eq.~\eqref{eq: quantum work} is minimized in the short-time limit by the $\delta-$STEP.

It is important to note the crucial assumption that the dissipator does not directly couple to position ($\mathcal{D}^{\dagger}g = 0$) is broken for some completely positive Markovian Lindblad dynamics~\cite{isar1994open}. In order to ensure positivity (and in some cases to ensure the uncertainty principle is not violated), an $\mathcal{O}(\hbar^2)$ correction is added to the dissipator which is neglected in the semi-classical limit of some quantum Brownian motion~\cite{isar1994open,el2026extracting}. If this dissipator is included, $\mathcal{O}(\Delta t)$ dynamics are restored for the averages of momentum-independent observables and delta-function kicks will no longer be optimal and instead the usual STEP result applies. 

Similar behavior was observed in Ref.~\cite{tokieda2025work} where the adiabatic Markovian master equation (A-GKSL) showed only small boundary jumps for a driven two-level system compared to the impulse-optimized protocols from the more accurate description of hierarchical equations of motion (HEOM), and the second-order time-convolutionless master equation (TCL2). Applying the short-time expansion to the dynamics governing these systems reveals that the A-GKSL has a linear in $\Delta t$ response even without impulses while an impulse is required for HEOM and TCL2 revealing the mechanism underlying this phenomena.

\section{Short-time expansion of the impulse work}
\label{App: Impulse work derivation}
Beginning with Eq.~\eqref{eq: impulse work}
\begin{align}
    \langle W^{\delta} \rangle = \frac{\langle\boldsymbol{A}^2\rangle_{\blambda_{0},\bp^+}}{2m} + \frac{\langle\boldsymbol{B}^2\rangle_{\Lambda}}{2m} -\frac{\langle\bp\cdot \boldsymbol{B}\rangle_{\Lambda}}{m} \ ,
    \label{eq: impulse work appendix}
\end{align}
a detailed derivation of Eq.~\eqref{eq: Kinetic work} follows by expanding the last two averages under the Klein--Kramers dynamics. To this end, we expand the final impulse as
\begin{align}
    \boldsymbol{B}(\br)=\boldsymbol{A}(\br)+\Delta t\boldsymbol{B}_{1}(\br)+\mathcal{O}(\Delta t^{2}) \ ,
    \label{eq: B expansion}
\end{align}
and recognize that for any phase-space function \(h(\br,\bp)\), its average immediately before the final impulse is
\begin{align}
    \langle h\rangle_{\Lambda}=\langle h\rangle_{\blambda_{0},\bp^{+}}+\Delta t\left\langle\mathcal{L}_{\blambda^{*}}^{\dagger}h\right\rangle_{\blambda_{0},\bp^{+}}+\mathcal{O}(\Delta t^{2}) \ ,
    \label{eq: short time adjoint expansion}
\end{align}
with adjoint Klein--Kramers operator
\begin{align}
    \mathcal{L}_{\blambda^{*}}^{\dagger}h=\frac{\bp}{m}\cdot\nabla_{\br}h+\left(\boldsymbol{F}_{\blambda^{*}}-\xi\bp\right)\cdot\nabla_{\bp}h+\frac{m\xi}{\beta}\nabla_{\bp}^{2}h \ .
    \label{eq: adjoint KK supplement}
\end{align}
Note that the averages are taken after the momentum is shifted according to
\begin{align}
    \bp^{+}=\bp+\boldsymbol{A}(\br) \ ,
\end{align}
and we have assumed a constant protocol $\blambda(t) = \blambda^*$.

We first consider \(\langle\boldsymbol{B}^{2}\rangle_{\Lambda}\). The explicit \(\Delta t\)-dependence of Eq.~\eqref{eq: B expansion} gives
\begin{align}
    \boldsymbol{B}^{2}=\boldsymbol{A}^{2}+2\Delta t\boldsymbol{A}\cdot\boldsymbol{B}_{1}+\mathcal{O}(\Delta t^{2}).
\end{align}
Since $\boldsymbol{A}^{2}$ is independent of momentum we can apply Eq.~\eqref{eq: short-time Average of p ind. variable} giving
\begin{align}
\langle\boldsymbol{B}^{2}\rangle_{\Lambda}=\langle\boldsymbol{A}^{2}\rangle_{\blambda_{0},\bp^{+}}+2\Delta t\langle\boldsymbol{A}\cdot\boldsymbol{B}_{1}\rangle_{\blambda_{0},\bp^{+}}+\frac{\Delta t}{m}\left\langle\bp\cdot\nabla_{\br}\boldsymbol{A}^{2}\right\rangle_{\blambda_{0},\bp^{+}}+\mathcal{O}(\Delta t^{2}) \ .
    \label{eq: B2 expansion}
\end{align}

To evaluate the final term, we define the pre-impulse momentum
\begin{align}
    \bq\equiv\bp-\boldsymbol{A}(\br) \ ,
\end{align}
under the shifted initial equilibrium Maxwell-Boltzmann distribution,
\begin{align}    \label{eq: shifted momentum moments}
    \langle\bq\rangle_{\blambda_{0},\bp^{+}}&=0 \ ,\\ \nonumber
    \langle\bq\otimes\bq\rangle_{\blambda_{0},\bp^{+}}&=\frac{m}{\beta}I \ ,
\end{align}
with $I$ the identity matrix. The pre-impulse momentum $\bq$ is independent of $\br$ implying
\begin{align}
    \langle\bp\rangle_{\blambda_{0},\bp^{+}}=\langle\boldsymbol{A}\rangle_{\blambda_{0},\bp^{+}} \ .
\end{align}
Using
\begin{align}
\bp\cdot\nabla_{\br}\boldsymbol{A}^{2}=2\boldsymbol{A}\cdot\left[(\bp\cdot\nabla_{\br})\boldsymbol{A}\right] \ ,
\end{align}
we obtain
\begin{align}
    \langle\boldsymbol{B}^{2}\rangle_{\Lambda}=\langle\boldsymbol{A}^{2}\rangle_{\blambda_{0}}+2\Delta t\langle\boldsymbol{A}\cdot\boldsymbol{B}_{1}\rangle_{\blambda_{0}}+\frac{2\Delta t}{m}\left\langle\boldsymbol{A}\cdot\left[(\boldsymbol{A}\cdot\nabla_{\br})\boldsymbol{A}\right]\right\rangle_{\blambda_{0}}+\mathcal{O}(\Delta t^{2}) \ .
    \label{eq: B2 final expansion}
\end{align}
Here and below, averages of quantities depending only on $\br$ are written as averages over the initial equilibrium distribution because the initial impulse does not change marginal distribution over $\br$.

We next consider $\langle\bp\cdot\boldsymbol{B}\rangle_{\Lambda}$. The explicit dependence of $\boldsymbol{B}$ on $\Delta t$ gives
\begin{align}
    \bp\cdot\boldsymbol{B}=\bp\cdot\boldsymbol{A}+\Delta t\bp\cdot\boldsymbol{B}_{1}+\mathcal{O}(\Delta t^{2}) \ .
\end{align}
Applying the adjoint generator to $\bp\cdot\boldsymbol{A}$ yields
\begin{align}
    \mathcal{L}_{\blambda^{*}}^{\dagger}\left(\bp\cdot\boldsymbol{A}\right)=\frac{1}{m}\bp\cdot\left[(\bp\cdot\nabla_{\br})\boldsymbol{A}\right]+\boldsymbol{F}_{\blambda^{*}}\cdot\boldsymbol{A}-\xi\bp\cdot\boldsymbol{A} \ .
    \label{eq: generator p dot A}
\end{align}
The momentum-diffusion term vanishes because $\nabla_{\bp}^{2}\left(\bp\cdot\boldsymbol{A}\right)=0$. Therefore, we have
\begin{align}
\langle\bp\cdot\boldsymbol{B}\rangle_{\Lambda}=\langle\bp\cdot\boldsymbol{A}\rangle_{\blambda_{0},\bp^{+}}+\Delta t\langle\bp\cdot\boldsymbol{B}_{1}\rangle_{\blambda_{0},\bp^{+}}+\Delta t\left\langle\frac{1}{m}\bp\cdot\left[(\bp\cdot\nabla_{\br})\boldsymbol{A}\right]+\boldsymbol{F}_{\blambda^{*}}\cdot\boldsymbol{A}-\xi\bp\cdot\boldsymbol{A}\right\rangle_{\blambda_{0},\bp^{+}}+\mathcal{O}(\Delta t^{2}) \ .
    \label{eq: p dot B intermediate}
\end{align}
The first two averages are
\begin{align}
\langle\bp\cdot\boldsymbol{A}\rangle_{\blambda_{0},\bp^{+}}=\langle\boldsymbol{A}^{2}\rangle_{\blambda_{0}} \ ,
\end{align}
and
\begin{align}
\langle\bp\cdot\boldsymbol{B}_{1}\rangle_{\blambda_{0},\bp^{+}}=\langle\boldsymbol{A}\cdot\boldsymbol{B}_{1}\rangle_{\blambda_{0}} \ .
\end{align}
The gradient term is evaluated using $\bp=\bq+\boldsymbol{A}$:
\begin{align}
    \left\langle\bp\cdot\left[(\bp\cdot\nabla_{\br})\boldsymbol{A}\right]\right\rangle_{\blambda_{0},\bp^{+}}=\left\langle\boldsymbol{A}\cdot\left[(\boldsymbol{A}\cdot\nabla_{\br})\boldsymbol{A}\right]\right\rangle_{\blambda_{0}}+\left\langle\bq\cdot\left[(\bq\cdot\nabla_{\br})\boldsymbol{A}\right]\right\rangle_{\blambda_{0},\bp^{+}}.
\end{align}
Using Eq.~\eqref{eq: shifted momentum moments}, the remaining contribution is
\begin{align}
    \left\langle\bq\cdot\left[(\bq\cdot\nabla_{\br})\boldsymbol{A}\right]\right\rangle_{\blambda_{0},\bp^{+}}=\frac{m}{\beta}\left\langle\nabla_{\br}\cdot\boldsymbol{A}\right\rangle_{\blambda_{0}}.
\end{align}
Therefore,
\begin{align}
\left\langle\bp\cdot\left[(\bp\cdot\nabla_{\br})\boldsymbol{A}\right]\right\rangle_{\blambda_{0},\bp^{+}}=\left\langle\boldsymbol{A}\cdot\left[(\boldsymbol{A}\cdot\nabla_{\br})\boldsymbol{A}\right]\right\rangle_{\blambda_{0}}+\frac{m}{\beta}\left\langle\nabla_{\br}\cdot\boldsymbol{A}\right\rangle_{\blambda_{0}}.
    \label{eq: momentum gradient average}
\end{align}
Substitution into Eq.~\eqref{eq: p dot B intermediate} gives
\begin{align}
    \langle\bp\cdot\boldsymbol{B}\rangle_{\Lambda}\approx\langle\boldsymbol{A}^{2}\rangle_{\blambda_{0}}+\Delta t\langle\boldsymbol{A}\cdot\boldsymbol{B}_{1}\rangle_{\blambda_{0}}+\Delta t\left\langle\boldsymbol{F}_{\blambda^{*}}\cdot\boldsymbol{A}-\xi\boldsymbol{A}^{2}\right\rangle_{\blambda_{0}}
    +\frac{\Delta t}{m}\left\langle\boldsymbol{A}\cdot\left[(\boldsymbol{A}\cdot\nabla_{\br})\boldsymbol{A}\right]\right\rangle_{\blambda_{0}}+\frac{\Delta t}{\beta}\left\langle\nabla_{\br}\cdot\boldsymbol{A}\right\rangle_{\blambda_{0}} \ .
    \label{eq: p dot B final expansion}
\end{align}

Substituting Eqs.~\eqref{eq: B2 final expansion} and \eqref{eq: p dot B final expansion} into Eq.~\eqref{eq: impulse work} gives the impulse work
\begin{align}
    \langle W^{\delta}\rangle_{\Lambda}\approx\frac{\Delta t}{m}\left[\xi\langle\boldsymbol{A}^{2}\rangle_{\blambda_{0}}-\left\langle\boldsymbol{A}\cdot\boldsymbol{F}_{\blambda^{*}}\right\rangle_{\blambda_{0}}-\frac{1}{\beta}\left\langle\nabla_{\br}\cdot\boldsymbol{A}\right\rangle_{\blambda_{0}}\right] \ ,
    \label{eq: impulse work before IBP}
\end{align}
and integration by parts applied to the final term yields
\begin{align}
    \frac{1}{\beta}\left\langle\nabla_{\br}\cdot\boldsymbol{A}\right\rangle_{\blambda_{0}}=-\left\langle\boldsymbol{A}\cdot\boldsymbol{F}_{\blambda_{0}}\right\rangle_{\blambda_{0}} \ .
    \label{eq: A divergence IBP}
\end{align}
Finally, substituting Eq.~\eqref{eq: A divergence IBP} into Eq.~\eqref{eq: impulse work before IBP} results in Eq.~\eqref{eq: Kinetic work}
\begin{align}
    \langle W^{\delta}\rangle_{\Lambda}\approx\frac{\Delta t}{m}\left\langle\xi\boldsymbol{A}^{2}-\boldsymbol{A}\cdot\left(\boldsymbol{F}_{\blambda^{*}}-\boldsymbol{F}_{\blambda_{0}}\right)\right\rangle_{\blambda_{0}} \ .
\end{align}

\section{Quadratic traps in arbitrary energy landscapes}\label{App: Quadratic}

Applying the general solution~\eqref{eq: Optimal short-time impulse} to a quadratic trapping potential in an arbitrary static landscape $V_{\rm land}(\br )$ we have
\begin{align}
    \boldsymbol{F} = K(t) (\br_{\rm trap}(t) - \br) - \nabla_{\br}V_{\rm land}(\br ) \ .
\end{align}
The force is not affine in $[K(t),\br^{\rm trap}(t)]$, but it is affine in $[K(t),\boldsymbol{q}(t)]$ with $\boldsymbol{q}(t)\equiv K(t)\br^{\rm trap}(t)$. Applying Eq.~\eqref{eq: Affine control} to the affine parameters we have
\begin{align}
    K^{\delta} &= \frac{\Delta K}{2\xi}\left[\delta(t) -\delta(t-\Delta t)\right] \ , \\
    \boldsymbol{q}^{\delta} &=  \frac{K_{\rm f}\br_{\rm trap, f}-K_{0}\br_{\rm trap,0}}{2\xi}\left[\delta(t) -\delta(t-\Delta t)\right] \ .
\end{align}
This impulse applies position dependent forces
\begin{align}    \label{eq: Optimal quadratic short-time impulse}
    &\boldsymbol{A}^* = \frac{1}{2\xi}\left[ K_{\rm f} (\br_{\rm trap, f} - \br) - K_0 (\br_{{\rm trap,}0} - \br) \right] \ ,
\end{align}
and $\boldsymbol{B}^* = \boldsymbol{A}^*+\mathcal{O}(\Delta t)$. For a stiffening trap, the impulse force is applied inwards $\boldsymbol{F}\propto-\br$ kicking the system towards a more confined state and for an expanding trap the force is applied outwards $\boldsymbol{F}\propto\br$.

When the stiffness is held fixed $K(t) = K$ the impulses become position independent. In this case the value for $\blambda^*$ is also particularly simple and exactly cancels out the $\mathcal{O}(\Delta t)$ correction to $\boldsymbol{B}^*$ in Eq.~\eqref{eq: Optimal short-time impulse}
\begin{align}\label{eq: Optimal quadratic trap centre short-time impulse}
    \br_{\rm trap}^* = \frac{\br_{\rm trap,f}+\br_{{\rm trap,}0}}{2} +\frac{\Delta\br_{\rm trap}}{2\xi}\left[\delta(t) -\delta(t-\Delta t)\right] \ .
\end{align}
This optimal protocol is in agreement with the original harmonic trapping potential protocol from Ref.~\cite{Gomez2008} taken to first order in $\Delta t$; however, our result is valid for arbitrary static energy landscapes while the original derivation is only valid for harmonic potentials in isolation.

For the protocol given in Eq.~\eqref{eq: Optimal quadratic trap centre short-time impulse}, the saved work is
\begin{align}
    \langle W_{\rm saved}^{\delta-{\rm STEP}} \rangle_{\Lambda} \approx \frac{k^2 \Delta\blambda^2\Delta t}{4m\xi} \ .
\end{align}
For comparison, the saved work from a linear protocol is
\begin{align}
    \langle W_{\rm saved}^{\rm linear} \rangle_{\Lambda} \approx \frac{k^2 \Delta\blambda^2\Delta t^2}{24m} \ .
\end{align}
This implies a ratio of optimal to linear of $6/(\xi\Delta t)$.


\begin{thebibliography}{51}%
	\makeatletter
	\providecommand \@ifxundefined [1]{%
		\@ifx{#1\undefined}
	}%
	\providecommand \@ifnum [1]{%
		\ifnum #1\expandafter \@firstoftwo
		\else \expandafter \@secondoftwo
		\fi
	}%
	\providecommand \@ifx [1]{%
		\ifx #1\expandafter \@firstoftwo
		\else \expandafter \@secondoftwo
		\fi
	}%
	\providecommand \natexlab [1]{#1}%
	\providecommand \enquote  [1]{``#1''}%
	\providecommand \bibnamefont  [1]{#1}%
	\providecommand \bibfnamefont [1]{#1}%
	\providecommand \citenamefont [1]{#1}%
	\providecommand \href@noop [0]{\@secondoftwo}%
	\providecommand \href [0]{\begingroup \@sanitize@url \@href}%
	\providecommand \@href[1]{\@@startlink{#1}\@@href}%
	\providecommand \@@href[1]{\endgroup#1\@@endlink}%
	\providecommand \@sanitize@url [0]{\catcode `\\12\catcode `\$12\catcode
		`\&12\catcode `\#12\catcode `\^12\catcode `\_12\catcode `\%12\relax}%
	\providecommand \@@startlink[1]{}%
	\providecommand \@@endlink[0]{}%
	\providecommand \url  [0]{\begingroup\@sanitize@url \@url }%
	\providecommand \@url [1]{\endgroup\@href {#1}{\urlprefix }}%
	\providecommand \urlprefix  [0]{URL }%
	\providecommand \Eprint [0]{\href }%
	\providecommand \doibase [0]{https://doi.org/}%
	\providecommand \selectlanguage [0]{\@gobble}%
	\providecommand \bibinfo  [0]{\@secondoftwo}%
	\providecommand \bibfield  [0]{\@secondoftwo}%
	\providecommand \translation [1]{[#1]}%
	\providecommand \BibitemOpen [0]{}%
	\providecommand \bibitemStop [0]{}%
	\providecommand \bibitemNoStop [0]{.\EOS\space}%
	\providecommand \EOS [0]{\spacefactor3000\relax}%
	\providecommand \BibitemShut  [1]{\csname bibitem#1\endcsname}%
	\let\auto@bib@innerbib\@empty
	\bibitem [{\citenamefont {Carnot}\ \emph {et~al.}(1960)\citenamefont {Carnot},
		\citenamefont {Clapeyron},\ and\ \citenamefont {Clausius}}]{Carnot1960}%
	\BibitemOpen
	\bibfield  {author} {\bibinfo {author} {\bibfnamefont {S.}~\bibnamefont
			{Carnot}}, \bibinfo {author} {\bibfnamefont {E.}~\bibnamefont {Clapeyron}},\
		and\ \bibinfo {author} {\bibfnamefont {R.}~\bibnamefont {Clausius}},\
	}\bibfield  {title} {\bibinfo {title} {Reflections on the motive power of
			fire},\ }in\ \href@noop {} {\emph {\bibinfo {booktitle} {Reflections on the
				Motive Power of Fire: And Others Papers on the Second Law of
				Thermodynamics}}},\ \bibinfo {editor} {edited by\ \bibinfo {editor}
		{\bibfnamefont {E.}~\bibnamefont {Mendoza}}}\ (\bibinfo  {publisher} {New
		York: Dover Publications},\ \bibinfo {year} {1960})\BibitemShut {NoStop}%
	\bibitem [{\citenamefont {Andresen}\ \emph {et~al.}(1977)\citenamefont
		{Andresen}, \citenamefont {Berry}, \citenamefont {Nitzan},\ and\
		\citenamefont {Salamon}}]{andresen1977}%
	\BibitemOpen
	\bibfield  {author} {\bibinfo {author} {\bibfnamefont {B.}~\bibnamefont
			{Andresen}}, \bibinfo {author} {\bibfnamefont {R.~S.}\ \bibnamefont {Berry}},
		\bibinfo {author} {\bibfnamefont {A.}~\bibnamefont {Nitzan}},\ and\ \bibinfo
		{author} {\bibfnamefont {P.}~\bibnamefont {Salamon}},\ }\bibfield  {title}
	{\bibinfo {title} {Thermodynamics in finite time. i. the step-carnot cycle},\
	}\href@noop {} {\bibfield  {journal} {\bibinfo  {journal} {Phys. Rev. A}\
		}\textbf {\bibinfo {volume} {15}},\ \bibinfo {pages} {2086} (\bibinfo {year}
		{1977})}\BibitemShut {NoStop}%
	\bibitem [{\citenamefont {Salamon}\ \emph {et~al.}(1977)\citenamefont
		{Salamon}, \citenamefont {Andresen},\ and\ \citenamefont
		{Berry}}]{salamon1977}%
	\BibitemOpen
	\bibfield  {author} {\bibinfo {author} {\bibfnamefont {P.}~\bibnamefont
			{Salamon}}, \bibinfo {author} {\bibfnamefont {B.}~\bibnamefont {Andresen}},\
		and\ \bibinfo {author} {\bibfnamefont {R.~S.}\ \bibnamefont {Berry}},\
	}\bibfield  {title} {\bibinfo {title} {Thermodynamics in finite time. ii.
			potentials for finite-time processes},\ }\href@noop {} {\bibfield  {journal}
		{\bibinfo  {journal} {Phys. Rev. A}\ }\textbf {\bibinfo {volume} {15}},\
		\bibinfo {pages} {2094} (\bibinfo {year} {1977})}\BibitemShut {NoStop}%
	\bibitem [{\citenamefont {Andresen}(2011)}]{andresen2011}%
	\BibitemOpen
	\bibfield  {author} {\bibinfo {author} {\bibfnamefont {B.}~\bibnamefont
			{Andresen}},\ }\bibfield  {title} {\bibinfo {title} {Current trends in
			finite-time thermodynamics},\ }\href@noop {} {\bibfield  {journal} {\bibinfo
			{journal} {Angew. Chem. Int. Ed.}\ }\textbf {\bibinfo {volume} {50}},\
		\bibinfo {pages} {2690} (\bibinfo {year} {2011})}\BibitemShut {NoStop}%
	\bibitem [{\citenamefont {Seifert}(2012)}]{Seifert2012}%
	\BibitemOpen
	\bibfield  {author} {\bibinfo {author} {\bibfnamefont {U.}~\bibnamefont
			{Seifert}},\ }\bibfield  {title} {\bibinfo {title} {Stochastic
			thermodynamics, fluctuation theorems and molecular machines},\ }\href@noop {}
	{\bibfield  {journal} {\bibinfo  {journal} {Rep. Prog. Phys.}\ }\textbf
		{\bibinfo {volume} {75}},\ \bibinfo {pages} {126001} (\bibinfo {year}
		{2012})}\BibitemShut {NoStop}%
	\bibitem [{\citenamefont {Brown}\ and\ \citenamefont
		{Sivak}(2017)}]{Brown2017}%
	\BibitemOpen
	\bibfield  {author} {\bibinfo {author} {\bibfnamefont {A.~I.}\ \bibnamefont
			{Brown}}\ and\ \bibinfo {author} {\bibfnamefont {D.~A.}\ \bibnamefont
			{Sivak}},\ }\bibfield  {title} {\bibinfo {title} {Toward the design
			principles of molecular machines},\ }\href@noop {} {\bibfield  {journal}
		{\bibinfo  {journal} {Physics in Canada}\ }\textbf {\bibinfo {volume} {73}}
		(\bibinfo {year} {2017})}\BibitemShut {NoStop}%
	\bibitem [{\citenamefont {Brown}\ and\ \citenamefont
		{Sivak}(2019)}]{Brown2019}%
	\BibitemOpen
	\bibfield  {author} {\bibinfo {author} {\bibfnamefont {A.~I.}\ \bibnamefont
			{Brown}}\ and\ \bibinfo {author} {\bibfnamefont {D.~A.}\ \bibnamefont
			{Sivak}},\ }\bibfield  {title} {\bibinfo {title} {Theory of nonequilibrium
			free energy transduction by molecular machines},\ }\href@noop {} {\bibfield
		{journal} {\bibinfo  {journal} {Chem. Rev.}\ }\textbf {\bibinfo {volume}
			{120}},\ \bibinfo {pages} {434} (\bibinfo {year} {2019})}\BibitemShut
	{NoStop}%
	\bibitem [{\citenamefont {Schmiedl}\ and\ \citenamefont
		{Seifert}(2007)}]{Schmiedl2007}%
	\BibitemOpen
	\bibfield  {author} {\bibinfo {author} {\bibfnamefont {T.}~\bibnamefont
			{Schmiedl}}\ and\ \bibinfo {author} {\bibfnamefont {U.}~\bibnamefont
			{Seifert}},\ }\bibfield  {title} {\bibinfo {title} {Optimal finite-time
			processes in stochastic thermodynamics},\ }\href@noop {} {\bibfield
		{journal} {\bibinfo  {journal} {Phys. Rev. Lett.}\ }\textbf {\bibinfo
			{volume} {98}},\ \bibinfo {pages} {108301} (\bibinfo {year}
		{2007})}\BibitemShut {NoStop}%
	\bibitem [{\citenamefont {Blaber}\ \emph {et~al.}(2021)\citenamefont {Blaber},
		\citenamefont {Louwerse},\ and\ \citenamefont {Sivak}}]{Blaber2021}%
	\BibitemOpen
	\bibfield  {author} {\bibinfo {author} {\bibfnamefont {S.}~\bibnamefont
			{Blaber}}, \bibinfo {author} {\bibfnamefont {M.~D.}\ \bibnamefont
			{Louwerse}},\ and\ \bibinfo {author} {\bibfnamefont {D.~A.}\ \bibnamefont
			{Sivak}},\ }\bibfield  {title} {\bibinfo {title} {Steps minimize dissipation
			in rapidly driven stochastic systems},\ }\href@noop {} {\bibfield  {journal}
		{\bibinfo  {journal} {Phys. Rev. E}\ }\textbf {\bibinfo {volume} {104}},\
		\bibinfo {pages} {L022101} (\bibinfo {year} {2021})}\BibitemShut {NoStop}%
	\bibitem [{\citenamefont {Rolandi}\ \emph {et~al.}(2023)\citenamefont
		{Rolandi}, \citenamefont {Perarnau-Llobet},\ and\ \citenamefont
		{Miller}}]{rolandi2023optimal}%
	\BibitemOpen
	\bibfield  {author} {\bibinfo {author} {\bibfnamefont {A.}~\bibnamefont
			{Rolandi}}, \bibinfo {author} {\bibfnamefont {M.}~\bibnamefont
			{Perarnau-Llobet}},\ and\ \bibinfo {author} {\bibfnamefont {H.~J.}\
			\bibnamefont {Miller}},\ }\bibfield  {title} {\bibinfo {title} {Optimal
			control of dissipation and work fluctuations for rapidly driven systems},\
	}\href@noop {} {\bibfield  {journal} {\bibinfo  {journal} {New Journal of
				Physics}\ }\textbf {\bibinfo {volume} {25}},\ \bibinfo {pages} {073005}
		(\bibinfo {year} {2023})}\BibitemShut {NoStop}%
	\bibitem [{\citenamefont {Gomez-Marin}\ \emph {et~al.}(2008)\citenamefont
		{Gomez-Marin}, \citenamefont {Schmiedl},\ and\ \citenamefont
		{Seifert}}]{Gomez2008}%
	\BibitemOpen
	\bibfield  {author} {\bibinfo {author} {\bibfnamefont {A.}~\bibnamefont
			{Gomez-Marin}}, \bibinfo {author} {\bibfnamefont {T.}~\bibnamefont
			{Schmiedl}},\ and\ \bibinfo {author} {\bibfnamefont {U.}~\bibnamefont
			{Seifert}},\ }\bibfield  {title} {\bibinfo {title} {Optimal protocols for
			minimal work processes in underdamped stochastic thermodynamics},\
	}\href@noop {} {\bibfield  {journal} {\bibinfo  {journal} {J. Chem. Phys.}\
		}\textbf {\bibinfo {volume} {129}},\ \bibinfo {pages} {024114} (\bibinfo
		{year} {2008})}\BibitemShut {NoStop}%
	\bibitem [{\citenamefont {Zhong}\ and\ \citenamefont
		{DeWeese}(2022)}]{zhong2022}%
	\BibitemOpen
	\bibfield  {author} {\bibinfo {author} {\bibfnamefont {A.}~\bibnamefont
			{Zhong}}\ and\ \bibinfo {author} {\bibfnamefont {M.~R.}\ \bibnamefont
			{DeWeese}},\ }\bibfield  {title} {\bibinfo {title} {Limited-control optimal
			protocols arbitrarily far from equilibrium},\ }\href@noop {} {\bibfield
		{journal} {\bibinfo  {journal} {Physical Review E}\ }\textbf {\bibinfo
			{volume} {106}},\ \bibinfo {pages} {044135} (\bibinfo {year}
		{2022})}\BibitemShut {NoStop}%
	\bibitem [{\citenamefont {Then}\ and\ \citenamefont {Engel}(2008)}]{Then2008}%
	\BibitemOpen
	\bibfield  {author} {\bibinfo {author} {\bibfnamefont {H.}~\bibnamefont
			{Then}}\ and\ \bibinfo {author} {\bibfnamefont {A.}~\bibnamefont {Engel}},\
	}\bibfield  {title} {\bibinfo {title} {Computing the optimal protocol for
			finite-time processes in stochastic thermodynamics},\ }\href@noop {}
	{\bibfield  {journal} {\bibinfo  {journal} {Phys. Rev. E}\ }\textbf {\bibinfo
			{volume} {77}},\ \bibinfo {pages} {041105} (\bibinfo {year}
		{2008})}\BibitemShut {NoStop}%
	\bibitem [{\citenamefont {Gingrich}\ \emph {et~al.}(2016)\citenamefont
		{Gingrich}, \citenamefont {Rotskoff}, \citenamefont {Crooks},\ and\
		\citenamefont {Geissler}}]{gingrich2016}%
	\BibitemOpen
	\bibfield  {author} {\bibinfo {author} {\bibfnamefont {T.~R.}\ \bibnamefont
			{Gingrich}}, \bibinfo {author} {\bibfnamefont {G.~M.}\ \bibnamefont
			{Rotskoff}}, \bibinfo {author} {\bibfnamefont {G.~E.}\ \bibnamefont
			{Crooks}},\ and\ \bibinfo {author} {\bibfnamefont {P.~L.}\ \bibnamefont
			{Geissler}},\ }\bibfield  {title} {\bibinfo {title} {Near-optimal protocols
			in complex nonequilibrium transformations},\ }\href@noop {} {\bibfield
		{journal} {\bibinfo  {journal} {Proc. Natl. Acad. Sci. U.S.A.}\ }\textbf
		{\bibinfo {volume} {113}},\ \bibinfo {pages} {10263} (\bibinfo {year}
		{2016})}\BibitemShut {NoStop}%
	\bibitem [{\citenamefont {Engel}\ \emph {et~al.}(2023)\citenamefont {Engel},
		\citenamefont {Smith},\ and\ \citenamefont {Brenner}}]{engel2022}%
	\BibitemOpen
	\bibfield  {author} {\bibinfo {author} {\bibfnamefont {M.~C.}\ \bibnamefont
			{Engel}}, \bibinfo {author} {\bibfnamefont {J.~A.}\ \bibnamefont {Smith}},\
		and\ \bibinfo {author} {\bibfnamefont {M.~P.}\ \bibnamefont {Brenner}},\
	}\bibfield  {title} {\bibinfo {title} {Optimal control of nonequilibrium
			systems through automatic differentiation},\ }\href@noop {} {\bibfield
		{journal} {\bibinfo  {journal} {Physical Review X}\ }\textbf {\bibinfo
			{volume} {13}},\ \bibinfo {pages} {041032} (\bibinfo {year}
		{2023})}\BibitemShut {NoStop}%
	\bibitem [{\citenamefont {Kamizaki}\ \emph {et~al.}(2022)\citenamefont
		{Kamizaki}, \citenamefont {Bonan{\c{c}}a},\ and\ \citenamefont
		{Muniz}}]{kamizaki2022}%
	\BibitemOpen
	\bibfield  {author} {\bibinfo {author} {\bibfnamefont {L.~P.}\ \bibnamefont
			{Kamizaki}}, \bibinfo {author} {\bibfnamefont {M.~V.}\ \bibnamefont
			{Bonan{\c{c}}a}},\ and\ \bibinfo {author} {\bibfnamefont {S.~R.}\
			\bibnamefont {Muniz}},\ }\bibfield  {title} {\bibinfo {title} {Performance of
			optimal linear-response processes in driven brownian motion far from
			equilibrium},\ }\href@noop {} {\bibfield  {journal} {\bibinfo  {journal}
			{Physical Review E}\ }\textbf {\bibinfo {volume} {106}},\ \bibinfo {pages}
		{064123} (\bibinfo {year} {2022})}\BibitemShut {NoStop}%
	\bibitem [{\citenamefont {Bonan{\c{c}}a}\ and\ \citenamefont
		{Deffner}(2018)}]{bonanca2018}%
	\BibitemOpen
	\bibfield  {author} {\bibinfo {author} {\bibfnamefont {M.~V.}\ \bibnamefont
			{Bonan{\c{c}}a}}\ and\ \bibinfo {author} {\bibfnamefont {S.}~\bibnamefont
			{Deffner}},\ }\bibfield  {title} {\bibinfo {title} {Minimal dissipation in
			processes far from equilibrium},\ }\href@noop {} {\bibfield  {journal}
		{\bibinfo  {journal} {Phys. Rev. E}\ }\textbf {\bibinfo {volume} {98}},\
		\bibinfo {pages} {042103} (\bibinfo {year} {2018})}\BibitemShut {NoStop}%
	\bibitem [{\citenamefont {Sivak}\ and\ \citenamefont
		{Crooks}(2016)}]{Sivak2016}%
	\BibitemOpen
	\bibfield  {author} {\bibinfo {author} {\bibfnamefont {D.~A.}\ \bibnamefont
			{Sivak}}\ and\ \bibinfo {author} {\bibfnamefont {G.~E.}\ \bibnamefont
			{Crooks}},\ }\bibfield  {title} {\bibinfo {title} {Thermodynamic geometry of
			minimum-dissipation driven barrier crossing},\ }\href@noop {} {\bibfield
		{journal} {\bibinfo  {journal} {Phys. Rev. E}\ }\textbf {\bibinfo {volume}
			{94}},\ \bibinfo {pages} {052106} (\bibinfo {year} {2016})}\BibitemShut
	{NoStop}%
	\bibitem [{\citenamefont {Blaber}\ and\ \citenamefont
		{Sivak}(2020{\natexlab{a}})}]{Blaber2020}%
	\BibitemOpen
	\bibfield  {author} {\bibinfo {author} {\bibfnamefont {S.}~\bibnamefont
			{Blaber}}\ and\ \bibinfo {author} {\bibfnamefont {D.~A.}\ \bibnamefont
			{Sivak}},\ }\bibfield  {title} {\bibinfo {title} {Optimal control of protein
			copy number},\ }\href@noop {} {\bibfield  {journal} {\bibinfo  {journal}
			{Phys. Rev. E}\ }\textbf {\bibinfo {volume} {101}},\ \bibinfo {pages}
		{022118} (\bibinfo {year} {2020}{\natexlab{a}})}\BibitemShut {NoStop}%
	\bibitem [{\citenamefont {Deffner}\ and\ \citenamefont
		{Bonan{\c{c}}a}(2020)}]{deffner2020}%
	\BibitemOpen
	\bibfield  {author} {\bibinfo {author} {\bibfnamefont {S.}~\bibnamefont
			{Deffner}}\ and\ \bibinfo {author} {\bibfnamefont {M.~V.}\ \bibnamefont
			{Bonan{\c{c}}a}},\ }\bibfield  {title} {\bibinfo {title} {Thermodynamic
			control—an old paradigm with new applications},\ }\href@noop {} {\bibfield
		{journal} {\bibinfo  {journal} {Europhys. Lett.}\ }\textbf {\bibinfo {volume}
			{131}},\ \bibinfo {pages} {20001} (\bibinfo {year} {2020})}\BibitemShut
	{NoStop}%
	\bibitem [{\citenamefont {Zulkowski}\ \emph {et~al.}(2012)\citenamefont
		{Zulkowski}, \citenamefont {Sivak}, \citenamefont {Crooks},\ and\
		\citenamefont {DeWeese}}]{zulkowski2012}%
	\BibitemOpen
	\bibfield  {author} {\bibinfo {author} {\bibfnamefont {P.~R.}\ \bibnamefont
			{Zulkowski}}, \bibinfo {author} {\bibfnamefont {D.~A.}\ \bibnamefont
			{Sivak}}, \bibinfo {author} {\bibfnamefont {G.~E.}\ \bibnamefont {Crooks}},\
		and\ \bibinfo {author} {\bibfnamefont {M.~R.}\ \bibnamefont {DeWeese}},\
	}\bibfield  {title} {\bibinfo {title} {Geometry of thermodynamic control},\
	}\href@noop {} {\bibfield  {journal} {\bibinfo  {journal} {Phys. Rev. E}\
		}\textbf {\bibinfo {volume} {86}},\ \bibinfo {pages} {041148} (\bibinfo
		{year} {2012})}\BibitemShut {NoStop}%
	\bibitem [{\citenamefont {Bonan{\c{c}}a}\ and\ \citenamefont
		{Deffner}(2014)}]{bonancca2014}%
	\BibitemOpen
	\bibfield  {author} {\bibinfo {author} {\bibfnamefont {M.~V.}\ \bibnamefont
			{Bonan{\c{c}}a}}\ and\ \bibinfo {author} {\bibfnamefont {S.}~\bibnamefont
			{Deffner}},\ }\bibfield  {title} {\bibinfo {title} {Optimal driving of
			isothermal processes close to equilibrium},\ }\href@noop {} {\bibfield
		{journal} {\bibinfo  {journal} {J. Chem. Phys.}\ }\textbf {\bibinfo {volume}
			{140}},\ \bibinfo {pages} {244119} (\bibinfo {year} {2014})}\BibitemShut
	{NoStop}%
	\bibitem [{\citenamefont {Zulkowski}\ and\ \citenamefont
		{DeWeese}(2015{\natexlab{a}})}]{zulkowski2015}%
	\BibitemOpen
	\bibfield  {author} {\bibinfo {author} {\bibfnamefont {P.~R.}\ \bibnamefont
			{Zulkowski}}\ and\ \bibinfo {author} {\bibfnamefont {M.~R.}\ \bibnamefont
			{DeWeese}},\ }\bibfield  {title} {\bibinfo {title} {Optimal control of
			overdamped systems},\ }\href@noop {} {\bibfield  {journal} {\bibinfo
			{journal} {Phys. Rev. E}\ }\textbf {\bibinfo {volume} {92}},\ \bibinfo
		{pages} {032117} (\bibinfo {year} {2015}{\natexlab{a}})}\BibitemShut
	{NoStop}%
	\bibitem [{\citenamefont {Zulkowski}\ and\ \citenamefont
		{DeWeese}(2015{\natexlab{b}})}]{zulkowski2015Quantum}%
	\BibitemOpen
	\bibfield  {author} {\bibinfo {author} {\bibfnamefont {P.~R.}\ \bibnamefont
			{Zulkowski}}\ and\ \bibinfo {author} {\bibfnamefont {M.~R.}\ \bibnamefont
			{DeWeese}},\ }\bibfield  {title} {\bibinfo {title} {Optimal protocols for
			slowly driven quantum systems},\ }\href@noop {} {\bibfield  {journal}
		{\bibinfo  {journal} {Phys. Rev. E}\ }\textbf {\bibinfo {volume} {92}},\
		\bibinfo {pages} {032113} (\bibinfo {year} {2015}{\natexlab{b}})}\BibitemShut
	{NoStop}%
	\bibitem [{\citenamefont {Large}\ and\ \citenamefont
		{Sivak}(2019)}]{Large2019}%
	\BibitemOpen
	\bibfield  {author} {\bibinfo {author} {\bibfnamefont {S.~J.}\ \bibnamefont
			{Large}}\ and\ \bibinfo {author} {\bibfnamefont {D.~A.}\ \bibnamefont
			{Sivak}},\ }\bibfield  {title} {\bibinfo {title} {Optimal discrete control:
			minimizing dissipation in discretely driven nonequilibrium systems},\
	}\href@noop {} {\bibfield  {journal} {\bibinfo  {journal} {J. Stat. Mech.:
				Theory Exp.}\ }\textbf {\bibinfo {volume} {2019}}\bibinfo  {number} { (8)},\
		\bibinfo {pages} {083212}}\BibitemShut {NoStop}%
	\bibitem [{\citenamefont {Lucero}\ \emph {et~al.}(2019)\citenamefont {Lucero},
		\citenamefont {Mehdizadeh},\ and\ \citenamefont {Sivak}}]{Lucero2019}%
	\BibitemOpen
	\bibfield  {number} {  }\bibfield  {author} {\bibinfo {author} {\bibfnamefont
			{J.~N.}\ \bibnamefont {Lucero}}, \bibinfo {author} {\bibfnamefont
			{A.}~\bibnamefont {Mehdizadeh}},\ and\ \bibinfo {author} {\bibfnamefont
			{D.~A.}\ \bibnamefont {Sivak}},\ }\bibfield  {title} {\bibinfo {title}
		{Optimal control of rotary motors},\ }\href@noop {} {\bibfield  {journal}
		{\bibinfo  {journal} {Phys. Rev. E}\ }\textbf {\bibinfo {volume} {99}},\
		\bibinfo {pages} {012119} (\bibinfo {year} {2019})}\BibitemShut {NoStop}%
	\bibitem [{\citenamefont {Rotskoff}\ and\ \citenamefont
		{Crooks}(2015)}]{Rotskoff2015}%
	\BibitemOpen
	\bibfield  {author} {\bibinfo {author} {\bibfnamefont {G.~M.}\ \bibnamefont
			{Rotskoff}}\ and\ \bibinfo {author} {\bibfnamefont {G.~E.}\ \bibnamefont
			{Crooks}},\ }\bibfield  {title} {\bibinfo {title} {Optimal control in
			nonequilibrium systems: Dynamic riemannian geometry of the ising model},\
	}\href@noop {} {\bibfield  {journal} {\bibinfo  {journal} {Phys. Rev. E}\
		}\textbf {\bibinfo {volume} {92}},\ \bibinfo {pages} {060102} (\bibinfo
		{year} {2015})}\BibitemShut {NoStop}%
	\bibitem [{\citenamefont {Rotskoff}\ \emph {et~al.}(2017)\citenamefont
		{Rotskoff}, \citenamefont {Crooks},\ and\ \citenamefont
		{Vanden-Eijnden}}]{Rotskoff2017}%
	\BibitemOpen
	\bibfield  {author} {\bibinfo {author} {\bibfnamefont {G.~M.}\ \bibnamefont
			{Rotskoff}}, \bibinfo {author} {\bibfnamefont {G.~E.}\ \bibnamefont
			{Crooks}},\ and\ \bibinfo {author} {\bibfnamefont {E.}~\bibnamefont
			{Vanden-Eijnden}},\ }\bibfield  {title} {\bibinfo {title} {Geometric approach
			to optimal nonequilibrium control: Minimizing dissipation in nanomagnetic
			spin systems},\ }\href@noop {} {\bibfield  {journal} {\bibinfo  {journal}
			{Phys. Rev. E}\ }\textbf {\bibinfo {volume} {95}},\ \bibinfo {pages} {012148}
		(\bibinfo {year} {2017})}\BibitemShut {NoStop}%
	\bibitem [{\citenamefont {Tafoya}\ \emph {et~al.}(2019)\citenamefont {Tafoya},
		\citenamefont {Large}, \citenamefont {Liu}, \citenamefont {Bustamante},\ and\
		\citenamefont {Sivak}}]{Tafoya2019}%
	\BibitemOpen
	\bibfield  {author} {\bibinfo {author} {\bibfnamefont {S.}~\bibnamefont
			{Tafoya}}, \bibinfo {author} {\bibfnamefont {S.~J.}\ \bibnamefont {Large}},
		\bibinfo {author} {\bibfnamefont {S.}~\bibnamefont {Liu}}, \bibinfo {author}
		{\bibfnamefont {C.}~\bibnamefont {Bustamante}},\ and\ \bibinfo {author}
		{\bibfnamefont {D.~A.}\ \bibnamefont {Sivak}},\ }\bibfield  {title} {\bibinfo
		{title} {Using a system’s equilibrium behavior to reduce its energy
			dissipation in nonequilibrium processes},\ }\href@noop {} {\bibfield
		{journal} {\bibinfo  {journal} {Proc. Natl. Acad. Sci. U.S.A}\ }\textbf
		{\bibinfo {volume} {116}},\ \bibinfo {pages} {5920} (\bibinfo {year}
		{2019})}\BibitemShut {NoStop}%
	\bibitem [{\citenamefont {Louwerse}\ and\ \citenamefont
		{Sivak}(2022)}]{louwerse2022}%
	\BibitemOpen
	\bibfield  {author} {\bibinfo {author} {\bibfnamefont {M.~D.}\ \bibnamefont
			{Louwerse}}\ and\ \bibinfo {author} {\bibfnamefont {D.~A.}\ \bibnamefont
			{Sivak}},\ }\bibfield  {title} {\bibinfo {title} {Multidimensional
			minimum-work control of a 2d ising model},\ }\href@noop {} {\bibfield
		{journal} {\bibinfo  {journal} {J. Chem. Phys}\ }\textbf {\bibinfo {volume}
			{156}},\ \bibinfo {pages} {194108} (\bibinfo {year} {2022})}\BibitemShut
	{NoStop}%
	\bibitem [{\citenamefont {Frim}\ and\ \citenamefont
		{DeWeese}(2022{\natexlab{a}})}]{frim2022}%
	\BibitemOpen
	\bibfield  {author} {\bibinfo {author} {\bibfnamefont {A.~G.}\ \bibnamefont
			{Frim}}\ and\ \bibinfo {author} {\bibfnamefont {M.~R.}\ \bibnamefont
			{DeWeese}},\ }\bibfield  {title} {\bibinfo {title} {Optimal finite-time
			brownian carnot engine},\ }\href@noop {} {\bibfield  {journal} {\bibinfo
			{journal} {Phys. Rev. E}\ }\textbf {\bibinfo {volume} {105}},\ \bibinfo
		{pages} {L052103} (\bibinfo {year} {2022}{\natexlab{a}})}\BibitemShut
	{NoStop}%
	\bibitem [{\citenamefont {Frim}\ and\ \citenamefont
		{DeWeese}(2022{\natexlab{b}})}]{frim2021}%
	\BibitemOpen
	\bibfield  {author} {\bibinfo {author} {\bibfnamefont {A.~G.}\ \bibnamefont
			{Frim}}\ and\ \bibinfo {author} {\bibfnamefont {M.~R.}\ \bibnamefont
			{DeWeese}},\ }\bibfield  {title} {\bibinfo {title} {Geometric bound on the
			efficiency of irreversible thermodynamic cycles},\ }\href@noop {} {\bibfield
		{journal} {\bibinfo  {journal} {Phys. Rev. Lett.}\ }\textbf {\bibinfo
			{volume} {128}},\ \bibinfo {pages} {230601} (\bibinfo {year}
		{2022}{\natexlab{b}})}\BibitemShut {NoStop}%
	\bibitem [{\citenamefont {Korbel}\ \emph {et~al.}(2026)\citenamefont {Korbel},
		\citenamefont {Kolchinsky}, \citenamefont {Loos}, \citenamefont {Manzano},
		\citenamefont {Garcia-Millan}, \citenamefont {Miangolarra},\ and\
		\citenamefont {Rold{\'a}n}}]{korbel2026quo}%
	\BibitemOpen
	\bibfield  {author} {\bibinfo {author} {\bibfnamefont {J.}~\bibnamefont
			{Korbel}}, \bibinfo {author} {\bibfnamefont {A.}~\bibnamefont {Kolchinsky}},
		\bibinfo {author} {\bibfnamefont {S.~A.}\ \bibnamefont {Loos}}, \bibinfo
		{author} {\bibfnamefont {G.}~\bibnamefont {Manzano}}, \bibinfo {author}
		{\bibfnamefont {R.}~\bibnamefont {Garcia-Millan}}, \bibinfo {author}
		{\bibfnamefont {O.~M.}\ \bibnamefont {Miangolarra}},\ and\ \bibinfo {author}
		{\bibfnamefont {{\'E}.}~\bibnamefont {Rold{\'a}n}},\ }\bibfield  {title}
	{\bibinfo {title} {Quo vadis, stochastic thermodynamics?},\ }\href@noop {}
	{\bibfield  {journal} {\bibinfo  {journal} {arXiv preprint arXiv:2604.26601}\
		} (\bibinfo {year} {2026})}\BibitemShut {NoStop}%
	\bibitem [{\citenamefont {Purcell}(2014)}]{purcell2014life}%
	\BibitemOpen
	\bibfield  {author} {\bibinfo {author} {\bibfnamefont {E.~M.}\ \bibnamefont
			{Purcell}},\ }\bibfield  {title} {\bibinfo {title} {Life at low reynolds
			number},\ }in\ \href@noop {} {\emph {\bibinfo {booktitle} {Physics and our
				world: reissue of the proceedings of a symposium in honor of Victor F
				Weisskopf}}}\ (\bibinfo {organization} {World Scientific},\ \bibinfo {year}
	{2014})\ pp.\ \bibinfo {pages} {47--67}\BibitemShut {NoStop}%
	\bibitem [{\citenamefont {Park}\ and\ \citenamefont
		{Schulten}(2004)}]{Park2004}%
	\BibitemOpen
	\bibfield  {author} {\bibinfo {author} {\bibfnamefont {S.}~\bibnamefont
			{Park}}\ and\ \bibinfo {author} {\bibfnamefont {K.}~\bibnamefont
			{Schulten}},\ }\bibfield  {title} {\bibinfo {title} {Calculating potentials
			of mean force from steered molecular dynamics simulations},\ }\href@noop {}
	{\bibfield  {journal} {\bibinfo  {journal} {J. Chem. Phys.}\ }\textbf
		{\bibinfo {volume} {120}},\ \bibinfo {pages} {5946} (\bibinfo {year}
		{2004})}\BibitemShut {NoStop}%
	\bibitem [{\citenamefont {Baldovin}\ \emph {et~al.}(2025)\citenamefont
		{Baldovin}, \citenamefont {Ben~Yedder}, \citenamefont {Plata}, \citenamefont
		{Raynal}, \citenamefont {Rondin}, \citenamefont {Trizac},\ and\ \citenamefont
		{Prados}}]{baldovin2025optimal}%
	\BibitemOpen
	\bibfield  {author} {\bibinfo {author} {\bibfnamefont {M.}~\bibnamefont
			{Baldovin}}, \bibinfo {author} {\bibfnamefont {I.}~\bibnamefont
			{Ben~Yedder}}, \bibinfo {author} {\bibfnamefont {C.~A.}\ \bibnamefont
			{Plata}}, \bibinfo {author} {\bibfnamefont {D.}~\bibnamefont {Raynal}},
		\bibinfo {author} {\bibfnamefont {L.}~\bibnamefont {Rondin}}, \bibinfo
		{author} {\bibfnamefont {E.}~\bibnamefont {Trizac}},\ and\ \bibinfo {author}
		{\bibfnamefont {A.}~\bibnamefont {Prados}},\ }\bibfield  {title} {\bibinfo
		{title} {Optimal control of levitated nanoparticles through finite-stiffness
			confinement},\ }\href@noop {} {\bibfield  {journal} {\bibinfo  {journal}
			{Physical Review Letters}\ }\textbf {\bibinfo {volume} {135}},\ \bibinfo
		{pages} {097102} (\bibinfo {year} {2025})}\BibitemShut {NoStop}%
	\bibitem [{\citenamefont {Rondin}\ \emph {et~al.}(2017)\citenamefont {Rondin},
		\citenamefont {Gieseler}, \citenamefont {Ricci}, \citenamefont {Quidant},
		\citenamefont {Dellago},\ and\ \citenamefont {Novotny}}]{rondin2017direct}%
	\BibitemOpen
	\bibfield  {author} {\bibinfo {author} {\bibfnamefont {L.}~\bibnamefont
			{Rondin}}, \bibinfo {author} {\bibfnamefont {J.}~\bibnamefont {Gieseler}},
		\bibinfo {author} {\bibfnamefont {F.}~\bibnamefont {Ricci}}, \bibinfo
		{author} {\bibfnamefont {R.}~\bibnamefont {Quidant}}, \bibinfo {author}
		{\bibfnamefont {C.}~\bibnamefont {Dellago}},\ and\ \bibinfo {author}
		{\bibfnamefont {L.}~\bibnamefont {Novotny}},\ }\bibfield  {title} {\bibinfo
		{title} {Direct measurement of kramers turnover with a levitated
			nanoparticle},\ }\href@noop {} {\bibfield  {journal} {\bibinfo  {journal}
			{Nature nanotechnology}\ }\textbf {\bibinfo {volume} {12}},\ \bibinfo {pages}
		{1130} (\bibinfo {year} {2017})}\BibitemShut {NoStop}%
	\bibitem [{\citenamefont {Militaru}\ \emph {et~al.}(2021)\citenamefont
		{Militaru}, \citenamefont {Lasanta}, \citenamefont {Frimmer}, \citenamefont
		{Bonilla}, \citenamefont {Novotny},\ and\ \citenamefont
		{Rica}}]{militaru2021kovacs}%
	\BibitemOpen
	\bibfield  {author} {\bibinfo {author} {\bibfnamefont {A.}~\bibnamefont
			{Militaru}}, \bibinfo {author} {\bibfnamefont {A.}~\bibnamefont {Lasanta}},
		\bibinfo {author} {\bibfnamefont {M.}~\bibnamefont {Frimmer}}, \bibinfo
		{author} {\bibfnamefont {L.~L.}\ \bibnamefont {Bonilla}}, \bibinfo {author}
		{\bibfnamefont {L.}~\bibnamefont {Novotny}},\ and\ \bibinfo {author}
		{\bibfnamefont {R.~A.}\ \bibnamefont {Rica}},\ }\bibfield  {title} {\bibinfo
		{title} {Kovacs memory effect with an optically levitated nanoparticle},\
	}\href@noop {} {\bibfield  {journal} {\bibinfo  {journal} {Physical Review
				Letters}\ }\textbf {\bibinfo {volume} {127}},\ \bibinfo {pages} {130603}
		(\bibinfo {year} {2021})}\BibitemShut {NoStop}%
	\bibitem [{\citenamefont {Dago}\ \emph {et~al.}(2021)\citenamefont {Dago},
		\citenamefont {Pereda}, \citenamefont {Barros}, \citenamefont {Ciliberto},\
		and\ \citenamefont {Bellon}}]{dago2021information}%
	\BibitemOpen
	\bibfield  {author} {\bibinfo {author} {\bibfnamefont {S.}~\bibnamefont
			{Dago}}, \bibinfo {author} {\bibfnamefont {J.}~\bibnamefont {Pereda}},
		\bibinfo {author} {\bibfnamefont {N.}~\bibnamefont {Barros}}, \bibinfo
		{author} {\bibfnamefont {S.}~\bibnamefont {Ciliberto}},\ and\ \bibinfo
		{author} {\bibfnamefont {L.}~\bibnamefont {Bellon}},\ }\bibfield  {title}
	{\bibinfo {title} {Information and thermodynamics: fast and precise approach
			to landauer’s bound in an underdamped micromechanical oscillator},\
	}\href@noop {} {\bibfield  {journal} {\bibinfo  {journal} {Physical Review
				Letters}\ }\textbf {\bibinfo {volume} {126}},\ \bibinfo {pages} {170601}
		(\bibinfo {year} {2021})}\BibitemShut {NoStop}%
	\bibitem [{\citenamefont {El~Qars}\ \emph {et~al.}(2026)\citenamefont
		{El~Qars}, \citenamefont {Kibbou}, \citenamefont {Essaoudi},\ and\
		\citenamefont {Ainane}}]{el2026extracting}%
	\BibitemOpen
	\bibfield  {author} {\bibinfo {author} {\bibfnamefont {J.}~\bibnamefont
			{El~Qars}}, \bibinfo {author} {\bibfnamefont {M.}~\bibnamefont {Kibbou}},
		\bibinfo {author} {\bibfnamefont {I.}~\bibnamefont {Essaoudi}},\ and\
		\bibinfo {author} {\bibfnamefont {A.}~\bibnamefont {Ainane}},\ }\bibfield
	{title} {\bibinfo {title} {Extracting work using a quantum optomechanical
			szilard-like engine},\ }\href@noop {} {\bibfield  {journal} {\bibinfo
			{journal} {Journal of Modern Optics}\ ,\ \bibinfo {pages} {1}} (\bibinfo
		{year} {2026})}\BibitemShut {NoStop}%
	\bibitem [{\citenamefont {Sabbagh}\ \emph {et~al.}(2024)\citenamefont
		{Sabbagh}, \citenamefont {Movilla~Miangolarra},\ and\ \citenamefont
		{Georgiou}}]{sabbagh2024wasserstein}%
	\BibitemOpen
	\bibfield  {author} {\bibinfo {author} {\bibfnamefont {R.}~\bibnamefont
			{Sabbagh}}, \bibinfo {author} {\bibfnamefont {O.}~\bibnamefont
			{Movilla~Miangolarra}},\ and\ \bibinfo {author} {\bibfnamefont {T.~T.}\
			\bibnamefont {Georgiou}},\ }\bibfield  {title} {\bibinfo {title} {Wasserstein
			speed limits for langevin systems},\ }\href@noop {} {\bibfield  {journal}
		{\bibinfo  {journal} {Physical Review Research}\ }\textbf {\bibinfo {volume}
			{6}},\ \bibinfo {pages} {033308} (\bibinfo {year} {2024})}\BibitemShut
	{NoStop}%
	\bibitem [{\citenamefont {Freitas}\ \emph {et~al.}(2020)\citenamefont
		{Freitas}, \citenamefont {Delvenne},\ and\ \citenamefont
		{Esposito}}]{freitas2020stochastic}%
	\BibitemOpen
	\bibfield  {author} {\bibinfo {author} {\bibfnamefont {N.}~\bibnamefont
			{Freitas}}, \bibinfo {author} {\bibfnamefont {J.-C.}\ \bibnamefont
			{Delvenne}},\ and\ \bibinfo {author} {\bibfnamefont {M.}~\bibnamefont
			{Esposito}},\ }\bibfield  {title} {\bibinfo {title} {Stochastic and quantum
			thermodynamics of driven rlc networks},\ }\href@noop {} {\bibfield  {journal}
		{\bibinfo  {journal} {Physical Review X}\ }\textbf {\bibinfo {volume} {10}},\
		\bibinfo {pages} {031005} (\bibinfo {year} {2020})}\BibitemShut {NoStop}%
	\bibitem [{\citenamefont {Tokieda}(2025)}]{tokieda2025work}%
	\BibitemOpen
	\bibfield  {author} {\bibinfo {author} {\bibfnamefont {M.}~\bibnamefont
			{Tokieda}},\ }\bibfield  {title} {\bibinfo {title} {Work-minimizing protocols
			in driven-dissipative quantum systems: An impulse-ansatz approach},\
	}\href@noop {} {\bibfield  {journal} {\bibinfo  {journal} {arXiv preprint
				arXiv:2511.15084}\ } (\bibinfo {year} {2025})}\BibitemShut {NoStop}%
	\bibitem [{\citenamefont {Shenfeld}\ \emph {et~al.}(2009)\citenamefont
		{Shenfeld}, \citenamefont {Xu}, \citenamefont {Eastwood}, \citenamefont
		{Dror},\ and\ \citenamefont {Shaw}}]{Shenfeld2009}%
	\BibitemOpen
	\bibfield  {author} {\bibinfo {author} {\bibfnamefont {D.~K.}\ \bibnamefont
			{Shenfeld}}, \bibinfo {author} {\bibfnamefont {H.}~\bibnamefont {Xu}},
		\bibinfo {author} {\bibfnamefont {M.~P.}\ \bibnamefont {Eastwood}}, \bibinfo
		{author} {\bibfnamefont {R.~O.}\ \bibnamefont {Dror}},\ and\ \bibinfo
		{author} {\bibfnamefont {D.~E.}\ \bibnamefont {Shaw}},\ }\bibfield  {title}
	{\bibinfo {title} {Minimizing thermodynamic length to select intermediate
			states for free-energy calculations and replica-exchange simulations},\
	}\href@noop {} {\bibfield  {journal} {\bibinfo  {journal} {Phys. Rev. E}\
		}\textbf {\bibinfo {volume} {80}},\ \bibinfo {pages} {046705} (\bibinfo
		{year} {2009})}\BibitemShut {NoStop}%
	\bibitem [{\citenamefont {Blaber}\ and\ \citenamefont
		{Sivak}(2020{\natexlab{b}})}]{Blaber2020Skewed}%
	\BibitemOpen
	\bibfield  {author} {\bibinfo {author} {\bibfnamefont {S.}~\bibnamefont
			{Blaber}}\ and\ \bibinfo {author} {\bibfnamefont {D.~A.}\ \bibnamefont
			{Sivak}},\ }\bibfield  {title} {\bibinfo {title} {Skewed thermodynamic
			geometry and optimal free energy estimation},\ }\href@noop {} {\bibfield
		{journal} {\bibinfo  {journal} {J. Chem. Phys.}\ }\textbf {\bibinfo {volume}
			{153}},\ \bibinfo {pages} {244119} (\bibinfo {year}
		{2020}{\natexlab{b}})}\BibitemShut {NoStop}%
	\bibitem [{\citenamefont {Ikeda}\ \emph {et~al.}(2025)\citenamefont {Ikeda},
		\citenamefont {Uda}, \citenamefont {Okanohara},\ and\ \citenamefont
		{Ito}}]{ikeda2025speed}%
	\BibitemOpen
	\bibfield  {author} {\bibinfo {author} {\bibfnamefont {K.}~\bibnamefont
			{Ikeda}}, \bibinfo {author} {\bibfnamefont {T.}~\bibnamefont {Uda}}, \bibinfo
		{author} {\bibfnamefont {D.}~\bibnamefont {Okanohara}},\ and\ \bibinfo
		{author} {\bibfnamefont {S.}~\bibnamefont {Ito}},\ }\bibfield  {title}
	{\bibinfo {title} {Speed-accuracy relations for diffusion models: Wisdom from
			nonequilibrium thermodynamics and optimal transport},\ }\href@noop {}
	{\bibfield  {journal} {\bibinfo  {journal} {Physical Review X}\ }\textbf
		{\bibinfo {volume} {15}},\ \bibinfo {pages} {031031} (\bibinfo {year}
		{2025})}\BibitemShut {NoStop}%
	\bibitem [{\citenamefont {Dockhorn}\ \emph {et~al.}(2022)\citenamefont
		{Dockhorn}, \citenamefont {Vahdat},\ and\ \citenamefont
		{Kreis}}]{dockhorn2021score}%
	\BibitemOpen
	\bibfield  {author} {\bibinfo {author} {\bibfnamefont {T.}~\bibnamefont
			{Dockhorn}}, \bibinfo {author} {\bibfnamefont {A.}~\bibnamefont {Vahdat}},\
		and\ \bibinfo {author} {\bibfnamefont {K.}~\bibnamefont {Kreis}},\ }\bibfield
	{title} {\bibinfo {title} {Score-based generative modeling with
			critically-damped langevin diffusion},\ }in\ \href
	{https://openreview.net/forum?id=CzceR82CYc} {\emph {\bibinfo {booktitle}
			{International Conference on Learning Representations}}}\ (\bibinfo {year}
	{2022})\BibitemShut {NoStop}%
	\bibitem [{\citenamefont {Blessing}\ \emph {et~al.}(2025)\citenamefont
		{Blessing}, \citenamefont {Berner}, \citenamefont {Richter},\ and\
		\citenamefont {Neumann}}]{blessing2025underdamped}%
	\BibitemOpen
	\bibfield  {author} {\bibinfo {author} {\bibfnamefont {D.}~\bibnamefont
			{Blessing}}, \bibinfo {author} {\bibfnamefont {J.}~\bibnamefont {Berner}},
		\bibinfo {author} {\bibfnamefont {L.}~\bibnamefont {Richter}},\ and\ \bibinfo
		{author} {\bibfnamefont {G.}~\bibnamefont {Neumann}},\ }\bibfield  {title}
	{\bibinfo {title} {Underdamped diffusion bridges with applications to
			sampling},\ }\href@noop {} {\bibfield  {journal} {\bibinfo  {journal} {arXiv
				preprint arXiv:2503.01006}\ } (\bibinfo {year} {2025})}\BibitemShut {NoStop}%
	\bibitem [{\citenamefont {Dechant}\ and\ \citenamefont
		{Sakurai}(2019)}]{dechant2019}%
	\BibitemOpen
	\bibfield  {author} {\bibinfo {author} {\bibfnamefont {A.}~\bibnamefont
			{Dechant}}\ and\ \bibinfo {author} {\bibfnamefont {Y.}~\bibnamefont
			{Sakurai}},\ }\bibfield  {title} {\bibinfo {title} {Thermodynamic
			interpretation of wasserstein distance},\ }\href@noop {} {\bibfield
		{journal} {\bibinfo  {journal} {arXiv preprint arXiv:1912.08405}\ } (\bibinfo
		{year} {2019})}\BibitemShut {NoStop}%
	\bibitem [{\citenamefont {Adelman}(1976)}]{adelman1976fokker}%
	\BibitemOpen
	\bibfield  {author} {\bibinfo {author} {\bibfnamefont {S.~A.}\ \bibnamefont
			{Adelman}},\ }\bibfield  {title} {\bibinfo {title} {Fokker--planck equations
			for simple non-markovian systems},\ }\href@noop {} {\bibfield  {journal}
		{\bibinfo  {journal} {The Journal of Chemical Physics}\ }\textbf {\bibinfo
			{volume} {64}},\ \bibinfo {pages} {124} (\bibinfo {year} {1976})}\BibitemShut
	{NoStop}%
	\bibitem [{\citenamefont {Isar}\ \emph {et~al.}(1994)\citenamefont {Isar},
		\citenamefont {Sandulescu}, \citenamefont {Scutaru}, \citenamefont
		{Stefanescu},\ and\ \citenamefont {Scheid}}]{isar1994open}%
	\BibitemOpen
	\bibfield  {author} {\bibinfo {author} {\bibfnamefont {A.}~\bibnamefont
			{Isar}}, \bibinfo {author} {\bibfnamefont {A.}~\bibnamefont {Sandulescu}},
		\bibinfo {author} {\bibfnamefont {H.}~\bibnamefont {Scutaru}}, \bibinfo
		{author} {\bibfnamefont {E.}~\bibnamefont {Stefanescu}},\ and\ \bibinfo
		{author} {\bibfnamefont {W.}~\bibnamefont {Scheid}},\ }\bibfield  {title}
	{\bibinfo {title} {Open quantum systems},\ }\href@noop {} {\bibfield
		{journal} {\bibinfo  {journal} {International Journal of Modern Physics E}\
		}\textbf {\bibinfo {volume} {3}},\ \bibinfo {pages} {635} (\bibinfo {year}
		{1994})}\BibitemShut {NoStop}%
\end{thebibliography}
\end{document}